\documentclass[journal,twocolumn]{IEEEtran}
\IEEEoverridecommandlockouts
\usepackage{amsmath}
\usepackage{amssymb}
\usepackage{amsfonts}
\usepackage{graphicx}
\usepackage{epsfig}
\usepackage{cite}
\usepackage{subfigure}
\usepackage{xcolor}
\usepackage[ruled,vlined,lined,ruled,linesnumbered]{algorithm2e}
\usepackage[font=small,skip=0pt]{caption}
\usepackage[belowskip=-15pt,aboveskip=0pt]{caption}
\usepackage{url}

\newcounter{mytempeqncnt}

\begin{document}

\title{Joint Optimization of Service Caching Placement and Computation Offloading in Mobile Edge Computing Systems}

\author{Suzhi~Bi, ~\IEEEmembership{Senior Member,~IEEE}, Liang~Huang, ~\IEEEmembership{Member,~IEEE}, and Ying-Jun Angela Zhang, ~\IEEEmembership{Fellow,~IEEE}\\
\thanks{S.~Bi is with the College of Electronic and Information Engineering, Shenzhen University, Shenzhen, Guangdong, China 518060 (e-mail: bsz@szu.edu.cn).}
\thanks{L.~Huang is with the College of Information Engineering, Zhejiang University of Technology, China (lianghuang@zjut.edu.cn).}
\thanks{Y-J.~A.~Zhang is with the Department of Information Engineering, The Chinese University of Hong Kong, Shatin, N.T., Hong Kong. (e-mail: yjzhang@ie.cuhk.edu.hk).}}
\maketitle

\begin{abstract}
In mobile edge computing (MEC) systems, edge service caching refers to pre-storing the necessary programs for executing computation tasks at MEC servers. Service caching effectively reduces the real-time delay/bandwidth cost on acquiring and initializing service applications when computation tasks are offloaded to the MEC servers. The limited caching space at resource-constrained edge servers calls for careful design of caching placement to determine which programs to cache over time. This is in general a complicated problem that highly correlates to the computation offloading decisions of computation tasks, i.e., whether or not to offload a task for edge execution. In this paper, we consider a single edge server that assists a mobile user (MU) in executing a sequence of computation tasks. In particular, the MU can upload and run its customized programs at the edge server, while the server can selectively cache the previously generated programs for future reuse. To minimize the computation delay and energy consumption of the MU, we formulate a mixed integer non-linear programming (MINLP) that jointly optimizes the service caching placement, computation offloading decisions, and system resource allocation (e.g., CPU processing frequency and transmit power of MU). To tackle the problem, we first derive the closed-form expressions of the optimal resource allocation solutions, and subsequently transform the MINLP into an equivalent pure 0-1 integer linear programming (ILP) that is much simpler to solve. To further reduce the complexity in solving the ILP, we exploit the underlying structures of caching causality and task dependency models, and accordingly devise a reduced-complexity alternating minimization technique to update the caching placement and offloading decision alternately. Extensive simulations show that the proposed joint optimization techniques achieve substantial resource savings of the MU compared to other representative benchmark methods considered.
\end{abstract}


\section{Introduction}
\subsection{Motivations and Summary of Contributions}
The proliferation of modern wireless applications, such as mobile gaming and augmented reality, demands persistent high-performance computations at commercial wireless devices to execute complex tasks with ultra-low latency. Over the last decade, large-scale \emph{cloud computing} platforms have been extensively deployed, which allows the wireless devices to offload intensive computation to remote cloud servers with abundant computing resource \cite{2010:Zhang}. To reduce the long backhaul transmission delay in the cloud, \emph{mobile edge computing} (MEC) has recently emerged to support ubiquitous high-performance computing, especially for delay-sensitive applications \cite{2017:Mao}. Specifically, MEC pushes publicly accessible computing resource to the edge of radio access network, e.g., cellular base stations and Wi-Fi access points, such that mobile users (MUs) can quickly offload computing tasks to their nearby edge servers.

Computing a task requires both the user task data as the input and the corresponding program that processes it. The use of \emph{caching} to dynamically store the program and/or task data at the MEC system has been recently recognized as a cost-effective method to reduce computation delay, energy consumption, and bandwidth cost. Here, we refer to caching the input and/or output data of computation tasks at the server/user side as \emph{computation content caching} (such as in \cite{2017:Cui,2018:Liu1,2018:Ko,2017:Lee,2019:Sun}). Likewise, we refer to caching the program data for executing computation tasks as \emph{computation service caching} (such as in \cite{2018:He,2019:Poularakis,2018:Xie,2018:Zhao,2018:Chen,2018:Xu}). Content caching reduces the frequency of repeated data transmissions and task computations. Its effectiveness relies on a strong assumption that the cached input/output data of a computation task is frequently reused by future executions. In practice, however, the input data and the corresponding computation output data of an application are rather dissimilar and hardly reusable for separate executions, e.g., human face recognition and interactive online gaming. In comparison, program data (and/or library data) in the cache is evidently reusable by future executions of the same application, e.g., the program and library for human face recognition. In an MEC platform, the cold start initialization of an application includes starting a cloud function, loading necessary libraries, and initializing user-specific code \cite{2019:Jonas}. Among them, it may take tens of seconds to load all the necessary application libraries. Caching the program data/library can effectively reduce the delay caused by application initialization or remote computation migration due to the absence of necessary program \cite{2010:Gunda}. Because edge servers are often limited in the caching space, a major design problem is to selectively cache service data over space (e.g., at multiple edge servers) and time for achieving optimum computing performance, e.g., minimum computation delay.

Existing work on mobile service caching has mostly assumed that the MU offloads all its computation tasks for edge/cloud execution. The tasks are executed at the edge server if the server has cached the required program. Otherwise, the edge server further offloads the task to a remote cloud server that can always compute the task at the cost of longer backhaul delay and larger bandwidth usage (e.g., see \cite{2018:He,2019:Poularakis,2018:Xie,2018:Zhao,2018:Chen,2018:Xu}). The focus of these works is to optimize the offline and online service caching placement decisions (i.e., what, when and where to cache) to minimize the computation workload forwarded to the cloud. Nonetheless, it is in general non-optimal to offload all computation tasks for edge/cloud execution. On one hand, the transmission of task data incurs long delay when the wireless channel condition is unfavorable. On the other hand, the edge computation may incur long delay in acquiring and initializing the service program when it is not pre-cached. Alternatively, \emph{opportunistic computation offloading} that allows the flexibility to execute some tasks locally at the MUs could be better off. Its performance advantage over full computation offloading has been verified by extensive recent studies under various network setups \cite{2017:You,2016:Chen,2018:Yan,2018:Wang,2018:Bi,2018:Liu,2016:Wang,2019:Alameddine,2019:Kherraf,2019:Hu,2019:Huang}. Notice that the task offloading decisions (i.e., whether offloading a task or computing locally) and the service caching placement are closely correlated. Intuitively, we tend to offload a task if the required program is already cached at the edge. Likewise, caching a program is cost-saving only if it is frequently reused to execute the tasks offloaded in the future. Therefore, it is necessary to jointly optimize service caching placement and offloading decisions in an MEC system. Such study, however, is largely overlooked in the existing literature.

Meanwhile, most existing work implicitly assumes that a central entity, e.g., the owner of the edge/cloud servers, is responsible for provisioning the program data in the cache, and that all required programs can be retrieved from a program pool in the backhaul network (e.g., in \cite{2018:He,2019:Poularakis,2018:Xie,2018:Zhao,2018:Chen,2018:Xu}). However, as the mobile computing scenarios become increasingly heterogeneous, it is common to allow the MUs to run custom-made or user-generated programs at the edge/cloud platforms. In fact, this is consistent with the concept of virtualization and infrastructure-as-a-service (IaaS) in edge/cloud computing paradigms, where the MUs are entitled to running their own programs using the physical resources of computing, storage, and networking provided by the infrastructure owners \cite{2014:Manvi}. For instance, \cite{2019:Hall} implemented an edge computing platform for image recognition using serverless functions. Multiple MUs send their own deep learning-based image recognition applications and personal images to an edge server, from which they receive the recognition results.\footnote{The image recognition application in \cite{2019:Hall} was written in C++ based on the tiny-dnn library (https://github.com/tiny-dnn/tiny-dnn), compiled to native x86 executable file in size 8.4 MB, and further executed within an OpenWhisk Apache container (http://openwhisk.apache.org/) at the edge server.} The application response latency is $744$ milliseconds at its initial call and is reduced to an average of $45$ milliseconds in the subsequent calls when the application service is cached. This implies that the overhead on uploading and initializing the application has significant impact to the service caching placement and offloading decisions.

In this paper, we consider an MEC system, where an edge server assists an MU in executing a sequence of $M$ \emph{dependent} tasks, where the output of one task is the input of the next one. Each task belongs to one of the $N$ applications and is either computed locally or offloaded for edge execution. In particular, the MU provides the program data for computing the tasks in the edge, while the edge server can selectively cache the previously generated programs and reuse them for processing future tasks. The detailed contributions of this paper are as follows.
\begin{itemize}
  \item We formulate a mixed integer non-linear programming (MINLP) problem to minimize the overall computation delay and energy consumption of the MU. Specifically, the problem jointly determines the optimal offloading decision of each task ($M$ binary variables), the service caching placement at the edge server throughout the task execution time ($MN$ binary variables), and system resource allocation (continuous variables representing the CPU processing frequency and transmit power of MU). The MINLP problem is in general lack of an efficient optimal algorithm in its original form.
  \item To tackle the problem, we first show that we can separately optimize the system resource allocation and derive the closed-form expressions of the optimal solutions. Based on the results, we then transform the MINLP into a pure 0-1 integer linear programming (ILP) problem that optimizes only the binary offloading decisions and service caching placements. The ILP problem can be handled by standard integer optimization algorithms, e.g., branch and bound method \cite{1998:Papadimitriou}. However, the exponential worst-case complexity can be high when either $M$ or $N$ is large.
  \item To gain more insight on the optimal solution structure and reduce the complexity of solving a large-size ILP problem, we first study the problem to optimize the $MN$ caching placement variables given the offloading decisions. By exploiting the caching causality property, we transform the original problem into a standard multidimensional knapsack problem (MKP). The MKP has no more than $M$ binary variables and can be efficiently handled by some off-the-shelf algorithms even if $M$ is relatively large, e.g., $M = 600$ \cite{2004:Freville}.
  \item We then optimize the $M$ offloading decisions given the caching placement. Interestingly, we find that the only difficulty lies in optimizing the offloading decisions of the ``uncached" tasks, whose required programs are not in the service cache. Meanwhile, the optimal offloading decisions of the remaining cached tasks can be easily retrieved. Together with our analysis on caching placement optimization, this leads to a reduced-complexity alternating minimization that iteratively updates the caching placements and offloading decisions.
\end{itemize}

Our simulations show that the joint optimization significantly reduces the computation delay and energy consumption of the MU compared to other benchmark methods. Meanwhile, the sub-optimal alternating minimization provides a reduced-complexity alternative for large-size problems. It is worth mentioning that this paper considers an offline model that assumes non-causal knowledge of future computation task parameters. The assumption is made to characterize the optimal structures of caching placement and offloading decisions. The obtained results can serve as an offline benchmark and may inspire future online algorithm designs that assume more practical prior knowledge.

\subsection{Related works}
Extensive prior work, e.g., \cite{2017:You,2016:Chen,2018:Yan,2018:Wang,2018:Bi,2018:Liu,2016:Wang,2019:Alameddine,2019:Kherraf,2019:Hu,2019:Huang}, has considered joint optimization of the task offloading decision (i.e., whether or how much data to offload) and system-level resource allocation (e.g., spectrum and computing power) to maximize the computational capability of an MEC system. Depending on the nature of the computation tasks, computation offloading is performed by following either a \emph{partial offloading} policy, i.e., an arbitrary part of the task data can be offloaded for edge execution, or a \emph{binary offloading} policy that an entire task is either offloaded or computed locally \cite{2017:Mao}. For instance, \cite{2017:You} optimizes a partial offloading policy in a multi-user MEC system to minimize the weighted sum energy consumption of the users. For multi-user MEC with binary offloading, \cite{2016:Chen} applies a separable semidefinite relaxation method to optimize the binary offloading decisions and wireless resource allocation. To support heterogenous computation tasks in IoT systems, \cite{2019:Alameddine} and \cite{2019:Kherraf} optimize the edge computation resource provisioning and task offloading/scheduling decisions to maximize the system operating efficiency (e.g., minimum operating cost or maximum accepted workload). To address the problem of high computation power consumption of wireless devices, \cite{2018:Wang,2018:Bi,2019:Huang} consider using wireless energy transfer technology to power wireless devices in MEC networks, and optimize the system computing performance under either partial or binary offloading policy. When the computation tasks at different MUs have input-output dependency, \cite{2018:Yan} studies the optimal binary offloading strategy and resource allocation to minimize the computation delay and energy consumption. The above work, however, ignores the use of edge caching to enhance the system-level computing performance.

On the other hand, recent work has applied content caching to MEC systems to effectively reduce computation delay, energy consumption, and bandwidth cost. In particular, an edge server can cache task output data \cite{2017:Cui}, task input data \cite{2018:Liu1}, and intermediate task computation results that are potentially useful for future task executions \cite{2017:Lee}. Meanwhile, content caching can also be implemented at the MU side to minimize the offloading (downloading) traffic to (from) the edge server \cite{2019:Sun}. To address the uncertainty of future task parameters, \cite{2017:Lee} and \cite{2018:Ko} propose online caching placement and prediction-based data prefetch methods. Despite their respective contributions, the fundamental assumption on reusing task input/output data may not hold for many mobile applications.

Computation service caching, on the other hand, caches the program data for processing a specific type of application. For instance, \cite{2018:He} considers caching program data of multiple applications in a set of collaborative BSs, and optimizing the caching placement and user-BS associations to minimize the data traffic forwarded to the remote cloud. A similar service caching placement problem is considered in \cite{2019:Poularakis} under communication, computation, and caching capacity constraints. Under the uncertainty of user service requests, e.g., application type and computation workload, \cite{2018:Xie} proposes a prediction-based online edge service caching algorithm to reduce the traffic load forwarded to the cloud. For a single edge server, \cite{2018:Zhao} assumes zero knowledge of future task arrivals and proposes an online service caching algorithm that achieves the optimal worst-case competitive ratio under homogeneous task arrivals. \cite{2018:Chen} proposes an online caching algorithm for collaborative edge servers to minimize the overall computation delay. Unlike \cite{2018:He,2019:Poularakis,2018:Xie,2018:Zhao,2018:Chen} that assume an entire task is computed either at an edge server or the cloud, \cite{2018:Xu} assumes that a task can be partitioned and executed in parallel at both the cloud and edge servers that have cached the necessary program.

All the above works neglect an important scenario where a task can be computed locally at the MU when edge execution is costly. Besides, they implicitly assume that a service program pool can provide all the programs required by the MUs. In this paper, we include local computation as an option for the MU, and allow the MU to upload its own programs to run at the edge server. In this case, the optimal caching placement is closely related to the offloading decisions, and vice versa, such that a joint optimization is required for maximum computation performance.

\begin{figure*}
\centering
  \begin{center}
    \includegraphics[width=0.7\textwidth]{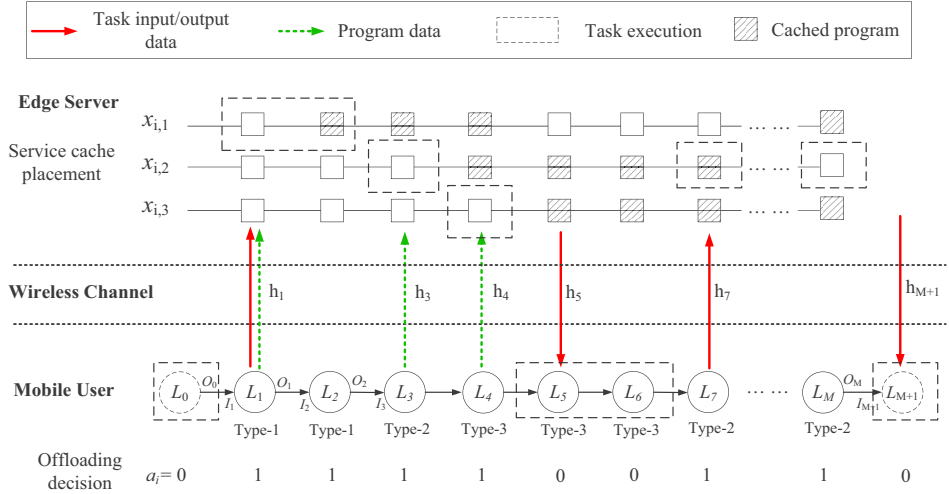}
  \end{center}
  \caption{Schematics of the considered service cache-assisted MEC system.}
  \label{model}
\end{figure*}

\section{System Model}
\subsection{Service Cache-assisted MEC System}
In Fig.~\ref{model}, we consider an edge server assisting the computation of an MU. This may correspond to a tagged MU in a multi-user network where each MU has been allocated with dedicated edge computing and communication resource to execute its own tasks. Some popular methods to achieve edge resource isolation include the container \cite{2017:Taleb} and serverless computing technologies \cite{2019:Jonas}, where the server resources (including CPU, memory, disk and networking resources, etc.) are partitioned into separate user space environments that do not interact with each other on the machine. For the tagged MU, we assume that it has a sequence of $M$ computation tasks to execute and each task is processed by one of the $N$ programs considered. We refer to a task as a type-$j$ task if it is processed by the $j$th program. Accordingly, we define a binary indicator $u_{i,j}$ such that $u_{i,j}=1$ if the $i$th task is a type-$j$ task, and $0$ otherwise. Besides, we denote the type of the $i$th task as $\varphi_i \in \left\{1,\cdots,N\right\}$, e.g., $\varphi_1 = 1$ and $\varphi_3 =2$ in Fig.~\ref{model}. The $M$ tasks are dependent such that the input of the $(i+1)$th task requires the output of the $i$th task, $i=1,\cdots, M-1$. One typical application of the sequential task execution model is eyeDentify \cite{2009:Kemp}, which consists of a series of steps for feature extraction operation in order to convert the raw image into a feature vector. Another application is a real time face recognition system in \cite{2014:Jaber}, which consists of a series of subtasks such as image extract, feature computation, histogram equalization, face detection, recognizer, and finding algorithm, etc. The optimal computation offloading problem in MEC system under the sequential task execution model have been studied in \cite{2015:Zhang} and \cite{2012:Zhang}, where the execution of each subtask only requires the output of the previous one.

The size of the input and output data of the $i$th task is denoted by $I_{i}$ and $O_i$, respectively. Besides, $L_i$ denotes the computing workload to process task $i$. For simplicity of illustration, we introduce two pseudo-tasks indexed as $0$ and $M+1$, and set $L_0 = L_{M+1} =0$, $O_{0} = I_{1}$ and $O_{M}=I_{M+1}$. Overall, the input and output task data sizes are related by $I_i = O_{i-1}, \ i=1,\cdots, M+1$. The MU follows the binary offloading policy so that each task can be computed either locally at the MU or offloaded to the edge server for remote execution. We use $a_i\in \left\{0,1\right\}$ to denote that the $i$th task is executed locally ($a_i=0$) or at the edge server ($a_i=1$). In particular, we set $a_0 =0$ and $a_{M+1}=0$, indicating that the series of computations initiate and terminate both at the MU. It is worth noting that the sequential task execution model can be extended to cascade the sub-tasks of multiple applications into a one ``super-task" consisting of all the sub-tasks.

Suppose that the MU runs its customized programs at the MEC platform by uploading its own program data (e.g., C/C++ code to generate a program). We denote the size of data to generate the $j$th program as $s_j$, $j=1,\cdots,N$. After receiving the program data, the edge server generates the corresponding program (e.g., the binary executable .EXE file) for processing the task data later offloaded. We denote the size of the $j$th generated program as $c_j$, where $c_j$ is in general much larger than $s_j$. Meanwhile, the edge server has a service cache that can cache the previously generated programs for future service reuse. We denote $x_{i,j}=1$ (or $0$) if the $j$th program is in the edge service cache (or not) before the execution of the $i$th task, either locally or at the edge, where $i=1,\cdots, M$. The edge server can decide to add (or remove) a program to (from) the cache during each task execution time, if the action is feasible under a finite caching space. For simplicity of illustration, we neglect the cost of adding or removing a program at the service cache and assume that the cache is empty initially, i.e., $x_{1,j}=0$ for all $j$.

Notice that the program data and task data can be offloaded separately. As an illustrative example in Fig.~\ref{model}, at the server side, a shaded (an empty) square in the $j$th row and $i$th column denotes $x_{i,j}=1$ ($x_{i,j}=0$). For example, the $1$st program is cached before the execution of the 2nd task until the end of the execution of the $4$th task (i.e., $x_{2,1}=x_{3,1}=x_{4,1}=1$). Besides, the dashed square denotes the location of task execution. For example, the $4$ tasks from the 1st until the 4th are executed at the edge, while the $5$th and $6$th tasks are executed locally (i.e., $a_1=a_2=a_3=a_4 =1$ and $a_5=a_6=0$ as shown in Fig.~\ref{model}). For the $1$st task to be executed at the edge, we need to upload both program and task data, as they are both absent at the server before the execution. However, we only need to offload the program data of the $3$rd and $4$th tasks, because their task input data is already at the edge as the output of previous edge computations. In addition, the $7$th task only uploads the task data, because the corresponding program is already in the edge service cache.

The detailed computation, caching, and communication models are described as follows:

\subsubsection{Computation Model} We assume that the MU has all the programs needed to process its tasks, e.g., pre-installed in the on-chip disk, such that the time consumed on processing a task $i$ locally only consists of the computation time.\footnote{For fast service data access and removal, edge servers cache the program data in high-speed memory, e.g., SRAM or RAM. Many commercial edge/cloud service platforms only allocate very limited memory to a user, e.g., the AWS Lambda platform provides the choice of 128 MBytes memory allocated to a user \cite{2019:Jonas}. In comparison, a common mobile phone has at least several Gigabytes disk memory available, which is slower but much less expensive, and sufficiently large to pre-install all the programs.} Specifically, the time and energy consumed on computing the $i$th task locally are \cite{2016:Wang}
\begin{equation}
\label{4}
\tau_i^l = \frac{L_i}{f_i}, \ \ e_i^l = \kappa f_i^\alpha \tau_i^l = \kappa \frac{(L_i)^\alpha}{(\tau_i^l)^{\alpha-1}},
\end{equation}
respectively, where $f_i$ denotes the local CPU frequency and is constrained by a maximum frequency $f_i \leq f_{max}$, $\kappa>0$ denotes the computing energy efficiency parameter, and $\alpha\geq 2$ denotes the exponent parameter.

When task $i$ is executed on the edge, the computation time includes two parts. First, the task processing time $\tau_i^c = \frac{L_i}{f_0}$, where $f_0$ denotes the fixed CPU frequency of the edge server. We assume that $f_0 > f_{max}$, i.e., the server has stronger computing power than the MU. Second, the server may need to generate a new program (e.g., program compilation and load function library) if it is not already in the cache. The program generation time of the $i$th task, if necessary, is $W_i \triangleq \sum_{j=1}^N u_{i,j} D_j$, where $D_j$ denotes the generation time of the $j$th program.

\subsubsection{Service Caching Model} We assume that the MU can only upload the program data for processing the current task that is executed at the edge. That is, the MU can only upload the $j$-th program data when executing the $i$th task at the edge if $u_{i,j}=1$ and $a_i=1$. Accordingly, $x_{i,j}=1$ is attainable only if at least one of the following two conditions holds:
\begin{enumerate}
  \item the $j$th program was in the cache before the execution of the last task ($x_{i-1,j}=1$);
  \item the $j$th program data was uploaded to the edge server in the last task execution time. This requires $u_{i-1,j}=1$ and $a_{i-1}=1$, or equivalently $a_{i-1} u_{i-1,j}=1$.
\end{enumerate}
If neither condition is satisfied, we have $x_{i,j}=0$. Equivalently, the above \emph{caching causality constraint} is expressed as
\begin{equation}
\label{101}
x_{i,j} \leq a_{i-1} u_{i-1,j}+ x_{i-1,j},
\end{equation}
for $i=1,\cdots, M, j=1,\cdots,N$. Besides, we need to observe the following caching capacity constraint throughout processing the $M$ tasks,
\begin{equation}
\label{100}
\mathsmaller \sum_{j=1}^N c_j\cdot x_{i,j} \leq C, \ i=1,\cdots,M,
\end{equation}
where $C$ is the caching space allocated by the MEC platform to serve the MU. In this paper, we assume that $C\geq \max_{j=1,\cdots,M}{c_j}$ to avoid trivial solution.

\subsubsection{Communication Model} Data transmissions between the edge server and the MU include uploading the program and/or task data, and downloading the computation result. For simplicity, we assume uplink/downlink channel reciprocity and use $h_i$ to denote the channel gain when transmitting the data of the $i$th task. We assume that $h_i$ remains constant during the data transmission of the $i$th task and may vary across different tasks. The uploading data rate for the $i$th task is $R^u_{i} = B \log_2\left(1 + \frac{p_i h_i}{\sigma_i^2}\right)$, where $B$ denotes the communication bandwidth, $p_i$ denotes the transmit power, and $\sigma_i^2$ denotes the receiver noise power including both potential interference and receiver thermal noise. Without loss of generality, we assume equal noise power for notation brevity, i.e., $\sigma_i^2=\sigma^2$ for $i = 1,\cdots,M$. Then, the time consumed on offloading the program data of the $i$th task is
\begin{equation}
\label{11}
\tau^s_i = \frac{\sum_{j=1}^N u_{i,j}s_{j}}{R^u_{i}} \triangleq  \frac{V_i}{R^u_{i}},
\end{equation}
where $V_i \triangleq \sum_{j=1}^N u_{i,j}s_j $ denotes the program data size of the $i$th task. Define function $g(x) = \sigma^2 \left(2^{\frac{x}{B}}-1\right)$. It follows from (\ref{11}) that the transmit power $p_i^s$ and the energy consumption $e_i^s$ are
\begin{equation}
\label{15}
p_i^s = \frac{1}{h_i} g\left(\frac{V_i }{\tau^s_i}\right),\ \ e^s_i = p_i^s \tau^s_i = \frac{\tau^s_i}{h_i} g\left(\frac{V_i }{\tau^s_i}\right),
\end{equation}
respectively. Notice that $e^s_i$ is convex with respect to $\tau^s_i$. Similarly, the time, power and energy spent on offloading the task data for the $i$th task are denoted as
\begin{equation}
\label{14}
\begin{aligned}
&\tau^u_i = \frac{O_{i-1}}{R^u_{i}}, \ \ p_i^u = \frac{1}{h_i} g\left(\frac{O_{i-1}}{\tau^u_i}\right), \\
&e^u_i = p_i^u \tau^u_i = \frac{\tau^u_i}{h_i} g\left(\frac{O_{i-1}}{\tau^u_i}\right),
\end{aligned}
\end{equation}
respectively. When both the task data and program data are offloaded to the edge, we assume that they are jointly encoded in one packet to reduce the packet header overhead. Accordingly, the edge server only starts initializing the program after receiving and decoding the whole packet. It can be easily verified that the time and energy consumed on transmitting both the program and task data of length $\left(V_i + O_{i-1}\right)$ are merely the sum of the corresponding two parts in (\ref{11})-(\ref{14}).

Furthermore, the time consumed on downloading the input data of the $i$th task for local computation is $\tau^d_i = \frac{O_{i-1}}{R^d_i}$, where $R^d_i = \bar{B} \log_2\left(1 + \frac{P_0 h_i}{\bar{\sigma}^2}\right)$ denotes a given downlink data rate for the $i$th task when the server transmits using fixed power $P_0$ and downlink bandwidth $\bar{B}$ under downlink receiver noise power $\bar{\sigma}^2$.

\subsection{Performance Metric}
The key performance metric considered in this paper is the total \emph{computation time and energy cost} (TEC) of the MU. In particular, the total computation time consists of two parts. The first part is the task execution time of the $M$ tasks, which can be expressed as
\begin{equation}
T^{exe} = \mathsmaller \sum_{i=1}^{M+1} \left[\left(1-a_i\right) \tau_i^l + a_i \tau_i^c \right].
\end{equation}
The two terms correspond to the processing delay that a task is executed locally and at edge server, respectively. The second part, denoted as $T^{pre}$, is the time spent on preparing for the program and task data before task execution, i.e., data transmission and program generation. Consider a tagged task $i$, we discuss the preparation time in the following cases.
\begin{enumerate}
  \item Case 1 ($a_{i-1}=0$ and $a_{i}=0$): In this case, the two consecutive tasks are computed locally, which incurs no delay on either program or task data transmission.
  \item Case 2 ($a_{i-1}=0$ and $a_{i}=1$): In this case, it takes $\tau^u_i$ amount of time to offload the task data to the edge. Meanwhile, program data uploading and program generation are needed if the program for computing the $i$th task is not in the cache. Mathematically, the delay overhead in offloading and initializing the program is
      \begin{equation}
       \tau_i^o \triangleq \left( W_i + \tau^s_i\right) \mathsmaller \sum_{j=1}^N (1-x_{i,j})u_{i,j}.
      \end{equation}
      Overall, the preparation time is $\tau^u_i + \tau_i^o$.
  \item Case 3 ($a_{i-1}=1$ and $a_{i}=0$): Only the computation output of the previous task needs to be downloaded to the MU. Accordingly, the time consumed is $\tau_i^d$.
  \item Case 4 ($a_{i-1}=1$ and $a_{i}=1$): The input task data of the $i$th task is already available after the computation of the previous task. Thus, the preparation time is the time needed for program data transmission and program generation, if the program data is not in the service cache. In other words, the time consumed is $\tau_i^o$.
\end{enumerate}
From the above analysis, we have
\begin{equation*}
\label{16}
T^{pre} = \mathsmaller \sum_{i=1}^{M+1} \left[ \left(1-a_{i-1}\right)a_{i} \tau^u_i + a_{i-1}\left(1-a_i\right)\tau^d_i + a_i \tau_i^o \right],
\end{equation*}
where $a_0 =a_{M+1} = 0$ by definition. Therefore, the total computation delay of the $M$ tasks is
\begin{equation}
\label{59}
\begin{aligned}
T &= T^{exe} + T^{pre} \\
&= \mathsmaller \sum_{i=1}^{M+1} \Big[ \left(1-a_{i-1}\right)a_{i} \tau^u_i + a_{i-1}\left(1-a_i\right)\tau^d_i \\
&\ \ \ \ \ \ \ \ \ \ \ \ \ + \left(1-a_i\right) \tau_i^l + a_i \tau_i^o + a_i \tau_i^c \Big].
\end{aligned}
\end{equation}

Meanwhile, the energy consumption of the MU is
\begin{equation}
\label{17}
E = \mathsmaller \sum_{i=1}^{M+1} \left[\left(1- a_i\right)e^l_i + \left(1-a_{i-1}\right)a_i e^u_i + a_i e^o_i\right],
\end{equation}
where $e^o_i = e^s_i \sum_{j=1}^N (1-x_{i,j}) u_{i,j}$ denotes the energy consumed on uploading the program data for the $i$th task. The other two terms correspond to the energy consumed on local computation and task data offloading, respectively. The performance metric TEC is the weighted sum of the two objectives, i.e., $TEC = \beta T + (1-\beta) E$, where $\beta \in [0,1]$ is a weighting parameter.

\section{Joint Caching Placement and Computation Offloading Optimization}
In this section, we formulate a joint optimization of resource allocation, caching placement and computation offloading decisions to minimize the TEC. We first derive the closed-form expressions of the optimal resource allocation. Accordingly, we show that (P2) can be equivalently transformed into a pure binary ILP problem, which can be handled by off-the-shelf algorithms.

\subsection{Problem Formulation}
In this paper, we are interested in minimizing the TEC of the MU by jointly optimizing the task offloading decision $\mathbf{a}$, the computational caching decision $\mathbf{X}$, and the system resource allocation $\left\{\mathbf{f},\boldsymbol{\tau},\mathbf{p}\right\}$. Here, $\mathbf{f} = \left\{f_i\right\}$, $\boldsymbol{\tau}=\left\{\tau_i^l, \tau_i^u,\tau_i^s\right\}$, $\boldsymbol{p}=\left\{p_i^u,p_i^s\right\}$. That is, we solve
 \begin{subequations}
   \begin{align}
    (\text{P1}):\  & \underset{\mathbf{a}, \mathbf{X},\mathbf{f},\boldsymbol{\tau},\mathbf{p}}{\text{minimize}} & &  \beta T + (1-\beta) E \label{61}\\
    & \text{subject to} &  & (\ref{101}), (\ref{100}), \label{68}\\
    & & & 0\leq p_i^u,p_i^s\leq P_{max}, \  \forall i,\label{65}\\
    & & & 0\leq f_i\leq f_{max},\ \forall i, \label{69}\\
    & & & \tau_i^l, \tau_i^u,\tau_i^s\geq 0, \forall i,\\
    & & & a_{i}, x_{i,j} \in \left\{0,1\right\},\  \forall i,j.
\end{align}
   \label{6}
\end{subequations}
Notice that $T$ and $E$ are non-linear functions of the optimizing variables, with the detailed expressions given in (\ref{59}) and (\ref{17}), respectively. (\ref{68}) corresponds to the caching causality and capacity constraints. (\ref{65}) and (\ref{69}) correspond to the maximum transmit power and CPU frequency of the MU. From (\ref{4}), there is a one-to-one mapping between $\tau_i^l$ and $f_i$. Besides, $p_i^u$ is uniquely determined by $\tau^u_i$ in (\ref{14}), and $p_i^s$ is uniquely determined by $\tau^s_i$ in (\ref{15}). By substituting $\left\{\mathbf{f},\mathbf{p}\right\}$ with $\boldsymbol{\tau}$, we can equivalently express (P1) as
 \begin{equation}
   \begin{aligned}
    (\text{P2}):\ & \underset{\mathbf{a}, \mathbf{X}, \boldsymbol{\tau}}{\text{minimize}} & &  \beta T + (1-\beta) E \label{61}\\
    & \text{subject to} &  &  (\ref{101}), (\ref{100}),\\
    & & & a_{i}, x_{i,j} \in \left\{0,1\right\},\ \forall i,j,\\
    & & & \tau_i^l \geq \frac{L_i}{f_{max}}, \ i=1,\cdots,M,\\
    & & & \tau_i^u \geq \frac{O_{i-1}}{R_i^{max}}, \ i=1,\cdots,M,\\
    & & & \tau_i^s \geq \frac{V_i}{R_i^{max}}, \ i=1,\cdots,M,
   \end{aligned}
\end{equation}
where $R_i^{max}=B\log_2\left(1+\frac{h_iP_{max}}{\sigma^2}\right)$ is a parameter. Problem (P2) is a mixed integer non-linear programming (MINLP), which lacks of efficient algorithm in its current form. In the following, we show that the problem can be equivalently transformed into a pure 0-1 integer linear programming (ILP).

\subsection{Optimal Resource Allocation}
A close observation of (P2) shows that the feasibility set of the binary variables $\left\{\mathbf{a},\mathbf{X}\right\}$ is not related to the resource allocation variables $\boldsymbol{\tau}$. Meanwhile, for any feasible $\left\{\mathbf{a},\mathbf{X}\right\}$, the optimal $\boldsymbol{\tau}^*$ is not related to $\left\{\mathbf{a},\mathbf{X}\right\}$, and thus can be separately optimized. Intuitively, this is because we can always decrease the objective by minimizing the energy and computation delay cost incurred by each task, regardless of the caching placement solution $\mathbf{X}$ and offloading decision $\mathbf{a}$. After plugging the optimal $\boldsymbol{\tau}^*$ back to (P2) and performing some simple manipulations, (P2) can be equivalently written as the following problem:
 \begin{equation}
   \begin{aligned}
   (P3):\ \ \ & \underset{\mathbf{a}, \mathbf{X}}{\text{minimize}} & & \mathsmaller \sum_{i=1}^{M+1} \rho_i \\
    & \text{subject to} &  & (\ref{101}), (\ref{100}), \ a_{i}, x_{i,j} \in \left\{0,1\right\},\ \forall i,j,
   \end{aligned}
\end{equation}
where
\begin{equation*}
\begin{aligned}
\rho_i \triangleq& o_i^* \left(1-a_{i-1}\right)a_i  + l_i^* \left(1-a_i\right) + s_i^* a_i \mathsmaller \sum_{j=1}^N \left(1-x_{i,j}\right) u_{i,j} \\
& + \beta a_{i-1}\left(1-a_i\right)\tau_i^d + \beta a_i \tau_i^c,
\end{aligned}
\end{equation*}
and $\left\{o_i^*,l_i^*,s_i^*\right\}$'s are parameters obtained by optimizing the resource allocation variables $\boldsymbol{\tau}$. Specifically, $o_i^*$ is obtained by optimizing $\tau_i^u$ as follows,
 \begin{subequations}
 \label{31}
   \begin{align}
    o_i^* = \ & \underset{\tau^u_i}{\text{minimize}} & &  \beta \tau_i^u + (1-\beta) \frac{\tau^u_i}{h_i} g\left(\frac{O_{i-1} }{\tau^u_i}\right)\\
    & \text{subject to} &  & \tau_i^u \geq \frac{O_{i-1}}{R_i^{max}},
   \end{align}
\end{subequations}
for $i=1,\cdots,M+1$. Likewise, $l_i^*$ is obtained by optimizing $\tau_i^l$ as follows,
 \begin{subequations}
 \label{32}
   \begin{align}
    l_i^* = \ & \underset{\tau^l_i}{\text{minimize}} & &  \beta \tau_i^l + (1-\beta)\kappa \frac{(L_i)^\alpha}{(\tau_i^l)^{\alpha-1}}\\
    & \text{subject to} &  & \tau_i^l \geq \frac{L_i}{f_{max}},
   \end{align}
\end{subequations}
for $i=1,\cdots,M+1$. In addition, $s_i^*$ is obtained by optimizing $\tau_i^s$ as follows,
 \begin{subequations}
 \label{33}
   \begin{align}
    s_i^* =\ & \underset{\tau_i^s}{\text{minimize}} & &  \beta  W_i + \beta\tau^s_i+ (1-\beta)\frac{\tau^s_i}{h_i} g\left(\frac{V_i }{\tau^s_i}\right)
    \\
    & \text{subject to} &  & \tau_i^s \geq \frac{V_i}{R_i^{max}},
   \end{align}
\end{subequations}
for $i=1,\cdots,M+1$. In other words, the optimization of the vector $\boldsymbol{\tau}$ can be decomposed into individual scalar optimization problems. The following Proposition 1 derives the closed-form expression of the optimal solution $\left(\tau^u_i\right)^*$ to (\ref{31}).

\textbf{Proposition 1:} The optimal solution $\tau_i^u$ is
\begin{equation}
\label{28}
\left(\tau_i^u\right)^* = \begin{cases}
\frac{O_{i-1}}{R_i^{max}},&   \text{if\ } h_i \leq \eta,\\
\frac{\ln 2 \cdot O_{i-1}}{B\cdot \left[  \mathcal{W}\left(e^{-1}\left[\frac{\beta h_i}{(1-\beta)\sigma^2} -1\right]\right)+ 1\right]},   &   \text{otherwise},\\
\end{cases}
\end{equation}
where $\eta \triangleq \frac{\sigma^2}{P_{max}}\left(\frac{A}{-\mathcal{W}\left( - A\exp\left(-A\right) \right)}-1\right)$ and $A = 1+ \frac{\beta}{(1-\beta)P_{max}}$ are fixed parameters. $\mathcal{W}(x)$ denotes the Lambert-W function, which is the inverse function of $J(z) = z \exp(z) =x$, i.e., $z = \mathcal{W}(x)$.

\emph{Proof:} Please see the detailed proof in Appendix A. $\hfill \blacksquare$

Because $\mathcal{W}(x) > -1$ when $x > -1/e$, the denominator in the second term of (\ref{28}) (and thus $\left(\tau_i^u\right)^*$) is always positive. Similar to the proof in Proposition 1, the optimal $\left(\tau_i^s\right)^*$ to (\ref{33}) is
\begin{equation}
\label{38}
\left(\tau_i^s\right)^* = \begin{cases}
\frac{V_i}{R_i^{max}},&   \text{if\ } h_i \leq \eta,\\
\frac{\ln 2 \cdot V_i}{B\cdot \left[  \mathcal{W}\left(e^{-1}\left[\frac{\beta h_i}{(1-\beta)\sigma^2} -1\right]\right)+ 1\right]},   &   \text{otherwise}.
\end{cases}
\end{equation}
Meanwhile, the optimal solution $\left(\tau_i^l\right)^*$ to (\ref{32}) can be obtained by calculating the derivative of the objective and considering the boundary condition, as follows
\begin{equation}
\label{39}
\left(\tau_i^l\right)^* = \begin{cases}
\frac{L_i}{f_{max}}, &   \text{if\ } f_{max} \leq  \left(\frac{\beta}{\kappa\left(1-\beta\right)\left(\alpha-1\right)}\right)^{\frac{1}{\alpha}}, \\
\left( \frac{ \kappa \left(1-\beta\right)\left(\alpha-1\right)}{\beta} \right)^{\frac{1}{\alpha}} L_i,   &   \text{otherwise}.\\
\end{cases}
\end{equation}
When the optimal $\boldsymbol{\tau}^*$ is obtained, the optimal $\left\{\mathbf{f}^*,\mathbf{p}^*\right\}$ in (P1) can be retrieved accordingly from (\ref{4}), (\ref{15}) and (\ref{14}).

\emph{Remark 1:} For the offloaded tasks, when $h_i$ is weaker than the fixed threshold in (\ref{28}), the MU should transmit at maximum power $\left(p_i^u\right)^* = P_{max}$ (or equivalently at the maximum data rate $R_i^{max}$) to minimize the offloading time. Otherwise, when $h_i$ is stronger than the threshold, the MU offloads for a shorter time $\left(\tau_i^u\right)^*$ when $h_i$ increases, because $\mathcal{W}\left(x\right)$ is an increasing function when $x > -1/e$. Similar results can also be obtained for $\left(\tau_i^s\right)^*$ and $\left(p_i^s\right)^*$ from (\ref{38}). For the tasks computed locally, the optimal solution $\left(\tau_i^l\right)^*$ in (\ref{39}) shows that the MU should compute faster either when a larger weight $\beta$ is assigned to the delay cost or when the local computation is more energy-efficient (small $\kappa$). When $\beta$ is sufficiently large or $\kappa$ is sufficiently small, the MU should compute at a maximum speed $f_{max}$ to minimize the computation delay.

\subsection{Equivalent ILP Formulation}\label{Joint}
Given the fixed parameters $\left\{o_i^*,l_i^*,s_i^*\right\}$ in (P3), the problem is a quadratic integer programming problem due to the multiplicative terms. To further simplify the problem, we introduce two sets of auxiliary variables $z_{i} \triangleq a_i x_{i,\varphi_{i}}$ and $b_i \triangleq a_ia_{i-1}$ for $i=1,\cdots,M$. Recall that $\varphi_{i}$ denotes the service type of the $i$th task. Accordingly, we re-express (P3) as
 \begin{subequations}
   \begin{align}
   & \underset{\mathbf{a,b,z,X}}{\text{minimize}} & & \sum_{i=1}^{M}  \left(o_i^* + \beta\tau_i^c + \beta \tau_{i+1}^d + s_i^* -l_i^* \right) a_i  \\
    & & & - \sum_{i=2}^{M} \left(o_i^* + \beta \tau_i^d\right)b_i  - \sum_{i=2}^{M} s_i^* z_{i}+ \sum_{i=1}^M l_i^*,\\
    & \text{subject to}   & &          b_i \leq  \frac{1}{2}\left(a_{i-1}+a_i\right), \ i=1,\cdots,M, \label{71} \\
    & & &          z_{i} \leq \frac{1}{2} \left(a_i + x_{i,\varphi_i}\right),\ i=1,\cdots, M, \label{72} \\
    & & &          (\ref{101}), (\ref{100}),\ a_{i}, b_i, z_{i},x_{i,j} \in \left\{0,1\right\},\ \forall i,j.
   \end{align}
   \label{76}
\end{subequations}
Constraint (\ref{71}) forces $b_i$ to be zero if either $a_{i-1}$ or $a_i$ is zero. Otherwise, if $a_{i-1} = a_i =1$, $b_i=1$ must hold at the optimum because the objective is decreasing in $b_i$. Therefore, $b_i = a_ia_{i-1}$ holds at the optimum when constraint (\ref{71}) is satisfied. Similar argument also applies to constraint (\ref{72}). Overall, the above problem is a standard 0-1 ILP problem, which can be handled by standard exact algorithms, e.g., branch and bound method \cite{1998:Papadimitriou}. Notice that the problem has $M(N+3)$ binary variables. The worst-case complexity of branch-and-bound method, as well as many other well-known exact algorithms for ILP, grows exponentially with the number of binary variables. Therefore, the complexity of solving (P3) can still be high when either $M$ or $N$ is large, e.g., taking several minutes to compute when $M$ equals several hundred. To reduce the complexity of solving a large-size ILP in real implementation, we investigate in the following sections an alternating minimization heuristic, where service caching placements and offloading decisions are optimized separately and iteratively.

\section{Optimal Service Caching Placement}
\subsection{Structure of the Caching Causality}\label{CacheOpt}
\begin{figure}
\centering
  \begin{center}
    \includegraphics[width=0.45\textwidth]{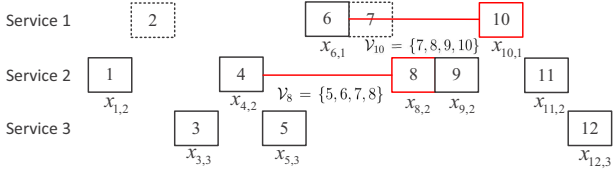}
  \end{center}
  \caption{An example task offloading decision. The rows represent the service types ($N=3$) and the columns represent the task indices ($M=12$). For instance, the $1$st task is type-$2$ and the $2$nd task is type-$1$. A solid square indicates the task is computed at the edge (i.e., $a_i=1$, such as task 6), and a dashed square indicates local computing (i.e., $a_i=0$, such as task 2). The variable below each solid box is the corresponding caching placement variable $x_{i,j}$. The red lines illustrate the index sets $\mathcal{V}_{8} = \{5,6,7,8\}$ and $\mathcal{V}_{10} = \{7,8,9,10\}$ in Definition 2.}
  \label{OffloadExample}
\end{figure}

In this section, we assume a feasible offloading decision $\mathbf{a}$ is given in (P3) and optimize the service caching placement $\mathbf{X}$. By eliminating the unrelated terms, (P3) reduces to
 \begin{subequations}
   \begin{align}
    (P4):\ \ \ & \underset{\mathbf{X}\in \left\{0,1\right\}^{M\times N}}{\text{maximize}} & &  \mathsmaller \sum_{i\in \mathcal{A}} s_i^* x_{i,\varphi_i}\\
    & \text{subject to} &  & (\ref{101}), (\ref{100}),
   \end{align}
\end{subequations}
where $\mathcal{A}$ denotes the index set of offloading tasks. For example, $\mathcal{A} = \{1,3,4,5,6,8,9,10,11,12\}$ in Fig.~\ref{OffloadExample}. There are in total $MN$ integer variables in (P4). However, as we show below, it is sufficient to optimize only the caching placement for the offloading tasks, i.e., $\left\{x_{i,\varphi_i}|i\in \mathcal{A}\right\}$, while the other variables are redundant. Let us first introduce the following two definitions to establish the caching causality of consecutive tasks.

\textbf{Definition 1:} Let $\nu_{i}^j =\left\{\min_{k\geq i} k |u_{k,j} = a_k = 1\right\}$ denote the index of the next type-$j$ task that is offloaded for edge execution since the execution of the $i$th task. In particular, $\nu_{i}^j =i$ if $u_{i,j} = a_i = 1$, and $\nu_{i}^j=\emptyset$ when no such task exists. For instance, $\nu_{11}^1 = \emptyset$, $\nu_{11}^2 = 11$, and $\nu_{11}^3 = 12$ in Fig.~\ref{OffloadExample}.

\textbf{Definition 2:} For a task $k \in \mathcal{A}$ of service type $\varphi_k$, we denote $\mathcal{V}_k = \left\{i \mid \nu_{i}^{\varphi_k} =k, i=1,\cdots,k\right\}$ as the index set of consecutive preceding tasks that satisfy $\nu_{i}^{\varphi_k} = k$. That is, $\nu_{i}^{\varphi_k} = k$ holds for $|\mathcal{V}_k|$ consecutive tasks from task $i=k- |\mathcal{V}_k| +1$ to task $k$, where $|\mathcal{V}_k|$ denotes the cardinality of $\mathcal{V}_k$. For instance, $\mathcal{V}_{8} = \{5,6,7,8\}$, $\mathcal{V}_{9} = \{9\}$, and $\mathcal{V}_{10} = \{7,8,9,10\}$ in Fig.~\ref{OffloadExample}.

We first show that we can safely set some variables $x_{i,j} =0$ without affecting the optimal value of (P4). Recall that $\nu_{1}^j$ denotes the index of the first type-$j$ task that is offloaded for edge execution (if any), e.g., $\nu_{1}^1 = 6$, $\nu_{1}^2 =1$, and $\nu_{1}^3=3$ in Fig.~\ref{OffloadExample}. By our assumption that the cache is initially empty, it holds from the caching causality constraint (\ref{101}) that $x_{\nu_{1}^j,j} =0$, $j=1,\cdots,N$. In Fig.~\ref{OffloadExample}, for instance, $x_{6,1}= x_{1,2}=x_{3,3} =0$. Besides, notice that for any $x_{i,j}$ satisfying $\nu_i^j = \emptyset$ (such as $x_{11,1}$), there is no type-$j$ task offloaded for edge execution afterwards. Thus, we can simply set $x_{l,j}=0$, for $l \geq i$, without affecting the optimal value of (P4). For instance, we can set $x_{11,1} = x_{12,1} = x_{12,2} = 0$ in Fig.~\ref{OffloadExample}. In this sense, we consider in the following only those variables $x_{i,j}$'s that satisfy $\nu_{i}^j \neq \emptyset$, i.e., $\left\{x_{i,\varphi_k}\mid \forall k\in \mathcal{A},\ \forall i \in \mathcal{V}_k\right\}$. The following Proposition $2$ proves that many of the remaining variables are redundant and can be removed from (P4).

\textbf{Proposition 2:} Suppose that $\mathbf{\hat{X}}=\{\hat{x}_{i,j}\}$ is a feasible solution of (P4), we can construct another feasible solution $\mathbf{\bar{X}}=\{\bar{x}_{i,j}\}$ by setting $\bar{x}_{i,\varphi_k} = \hat{x}_{k,\varphi_k}, \forall k\in \mathcal{A} \text{ and } \forall i \in \mathcal{V}_k$. Meanwhile, the objective values of (P4) are the same with $\mathbf{\hat{X}}$ and $\mathbf{\bar{X}}$.

\emph{Proof:} Please see the detailed proof in Appendix B. $\hfill \blacksquare$

\emph{Remark 2}: Proposition $2$ indeed shows that only $\left\{x_{k,\varphi_k} \mid \forall k\in \mathcal{A}\right\}$ are independent variables, while the rest of the variables $\{x_{i,j}\}$ in (P4) are dependent variables and redundant. For a tagged offloading task $k \in \mathcal{A}$, by replacing $x_{i,\varphi_k}$ with $x_{k,\varphi_k}, \forall i \in \mathcal{V}_k$ in (P4), we remove not only the dependent variables in $\mathcal{V}_k$, but also the redundant constraints in (\ref{101}) for $i \in \mathcal{V}_k$ and $j= \varphi_k$. Furthermore, by considering all $k \in \mathcal{A}$, we remove all dependent variables and all the constraints in (\ref{101}) without affecting the optimal value of (P4).

Take Fig.~\ref{OffloadExample} as an example. We can intuitively visualize the variable replacements derived from Proposition $2$. That is, each caching placement variable $x_{i,j}$ corresponding to a blank slot (e.g., $x_{7,2}$) or a dashed square (e.g., $x_{7,1}$) can be equivalently replaced by that corresponding to the next solid square in the same row (e.g., by $x_{8,2}$ and $x_{10,1}$, respectively).  For instance, we can replace $\left\{x_{7,1}, x_{8,1}, x_{9,1}\right\}$ by $x_{10,1}$ while achieving the same optimal value of (P4). The replacement also removes the constraints in (\ref{101}) for $i=7,8,9,10$ and $j=1$.

Following the variable replacement technique proposed in Proposition $2$, (P4) is equivalently transformed to the following problem
 \begin{subequations}
   \begin{align}
    (\text{P4-Eq}):\ & \underset{x_{i,\varphi_i}, \forall i \in  \mathcal{A}}{\text{maximize}} & &  \mathsmaller \sum_{i\in \mathcal{A}} s_i^* x_{i,\varphi_i}\\
    & \text{subject to} &  & x_{i,\varphi_i}\in \left\{0,1\right\},\ \forall i \in  \mathcal{A},\\
    & & &                    x_{\nu_{1}^j,j} =0, \ j=1,\cdots,N, \\
    & & &                    \mathsmaller \sum_{j=1}^N c_j \cdot x_{\nu_{i}^j,j} \leq C, \ \forall i. \label{43}
   \end{align}
\end{subequations}
Notice that the above problem (P4-Eq) is a standard multidimensional knapsack problem (MKP) \cite{2004:Freville}. Compared to (P4), the number of binary variables is reduced from $MN$ to only $|\mathcal{A}|\leq M$. Besides, as we will illustrate in the next subsection, many constraints in (\ref{43}) are duplicated or redundant. When there is more than one effective constraint in (P4-Eq), there does not exist a fully polynomial-time approximation scheme (FPTAS) \cite{2004:Freville}. However, for MKP problems of moderate size, plenty of algorithms include hybrid dynamic programming and branch-and-bound methods can be applied to solve for the exact optimal solution in a reasonable computation time, e.g., within $0.1$ second of computation time for over $500$ variables \cite{2013:Berger}.

\subsection{Optimal Caching Placement: A Case Study}
In this subsection, we use the example in Fig. \ref{OffloadExample} to illustrate the problem transformation from (P4) to (P4-Eq). We first apply the above variable replacement technique to the $M$ constraints in (\ref{100}) of (P4) one by one to construct the corresponding $M$ constraints in (\ref{43}) of (P4-Eq). Starting from the first constraint in (\ref{43}), we note that $\left\{\nu_{1}^1, \nu_{1}^2,\nu_{1}^3\right\} = \left\{6,1,3\right\}$, and thus focus on variables $\{x_{6,1}, x_{1,2}, x_{3,3}\}$. Because the service cache is assumed empty initially, we have $x_{6,1}=x_{1,2}=x_{3,3}=0$. Therefore, the first constraint in (\ref{43}) of (P4-Eq) is unnecessary. The second constraint in (\ref{43}) can be expressed as $C_2:  c_2 x_{4,2} \leq C$ because $\nu_{2}^2 = 4$. After applying the similar variable replacement procedure to constraint $i=3,\cdots,12$, we obtain the $M$ constraints of (P4-Eq), which however contain duplicated or redundant constraints. For instance, $C_2$ is redundant due to the assumption that $C\geq c_i$, $\forall i$. Meanwhile, it can be easily verified that the 3rd constraint $C_3$ in (\ref{43}) is the same as $C_2$. In addition, for the $6$th and $7$th constraints in (\ref{43}), we have
\begin{equation*}
\begin{aligned}
&C_6: c_2 x_{8,2} + c_3 x_{12,3} \leq C,\\
&C_7: c_1 x_{10,1} + c_2 x_{8,2} + c_3 x_{12,3} \leq C,
\end{aligned}
\end{equation*}
where $C_6$ is evidently redundant if $C_7$ is satisfied.

After removing all the duplicated/redundant constraints, (P4-Eq) becomes
 \begin{equation}
 \label{44}
 \begin{aligned}
&\underset{x_{i,\varphi_i},\forall i \in  \bar{\mathcal{A}}}{\text{maximize}} & &   \sum_{i\in \bar{\mathcal{A}}} s_i^* x_{i,\varphi_i}\\
& \text{subject to}  & &  x_{i,\varphi_i}\in \left\{0,1\right\},\ \forall i \in  \bar{\mathcal{A}},\\
& & &                     C_4: c_2 x_{4,2} + c_3 x_{5,3} \leq C,\\
& & &                     C_5: c_2 x_{8,2} + c_3 x_{5,3} \leq C, \\
& & &                     C_7: c_1 x_{10,1} + c_2 x_{8,2} + c_3 x_{12,3} \leq C, \\
& & &                     C_9: c_1 x_{10,1} + c_2 x_{9,2} + c_3 x_{12,3} \leq C,\\
& & &                     C_{10}: c_1 x_{10,1} + c_2 x_{11,2} + c_3 x_{12,3}\leq C,
\end{aligned}
\end{equation}
where $\bar{\mathcal{A}} \triangleq \left\{4,5,8,9,10,11,12\right\}$ denotes the indices of the remaining offloading tasks. Compared with its original formulation in (P4), the numbers of binary variables are reduced from $MN=36$ to $7$, the number of caching capacity constraints is reduced from $M= 12$ to $5$, and all the caching causality constraints are removed. Besides, the original generic ILP is converted to a standard 0-1 MKP, where many specialized exact and approximate solution algorithms are available.

\begin{figure}
\centering
  \begin{center}
    \includegraphics[width=0.47\textwidth]{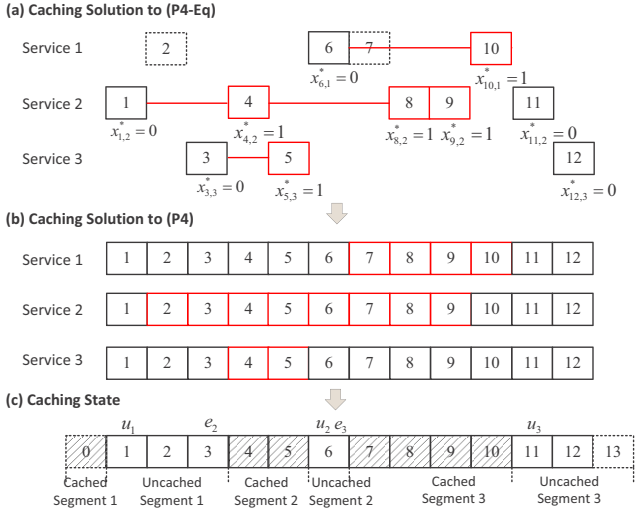}
  \end{center}
  \caption{(a). An example optimal solution to (P4-Eq) adapted from Fig.~\ref{OffloadExample}, where a red (black) box indicates $x^*_{i,j}=1$ ($0$); (b). The retrieved caching solution to (P4); (c). The caching state derived from the caching solution. A shaded (empty) box indicates $x_{i,\varphi_i}=1$ ($0$).}
  \label{CacheExample}
\end{figure}

After solving (\ref{44}) optimally, we can easily retrieve the solution in (P4) by Proposition $2$. For example, for the $2$nd program, the optimal solutions can be retrieved from $\left\{ x^*_{4,2}, x^*_{8,2}, x^*_{9,2}, x^*_{11,2}\right\}$ as $x^*_{2,2} =x^*_{3,2}= x^*_{4,2}$, $x^*_{5,2} = x^*_{6,2} = x^*_{7,2} = x^*_{8,2}$, $x^*_{10,2} = x^*_{11,2}$, while $x^*_{i,2} = 0$ for the rest task $i$. The optimal solution of $x^*_{i,1}$'s and $x^*_{i,3}$'s can be similarly obtained. As an illustrating example, suppose that Fig. \ref{CacheExample}(a) is the optimal caching solution to (P4-Eq) for the example in Fig. \ref{OffloadExample}, Fig. \ref{CacheExample}(b) shows the corresponding optimal caching placement solution to (P4).

\section{Optimal Task Offloading Decision}
In this section, we optimize the task offloading decision $\mathbf{a}$ given a caching placement decision $\mathbf{X}$ in (P3). Interestingly, we find that the only difficulty lies in optimizing the offloading decisions of the ``uncached" tasks, which effectively reduces the number of binary variables.

\subsection{Reduced-Complexity Decomposition Method}\label{decomp}
Notice that once $\mathbf{X}$ is given, (P3) is reduced to
 \begin{subequations}
 \label{73}
   \begin{align}
    \ \ \ & \underset{\mathbf{a}}{\text{minimize}} & & \mathsmaller \sum_{i=1}^{M+1}\big\{ o_i^* \left(1-a_{i-1}\right)a_i  + l_i^* \left(1-a_i\right) \\
    & & & + \left(\lambda_i^*+ \beta \tau_i^c\right) a_i   + \beta \tau_i^d a_{i-1}\left(1-a_i\right) \big\}\\
    & \text{subject to}  & &  u_{i-1,j} a_{i-1}  \geq x_{i,j} - x_{i-1,j}, \ \forall i,j, \label{53}\\
    & & & a_{i} \in \left\{0,1\right\},\ i=1,\cdots,M,
   \end{align}
\end{subequations}
where $a_0 = 0$ and $a_{M+1}=0$. Here, $\lambda_i^*$ in the objective is a parameter determined by the value of $x_{i,\varphi_i}$, where $\lambda_i^* = 0$ if $x_{i,\varphi_i}=1$ and $\lambda_i^* = s_i^*$ if $x_{i,\varphi_i}=0$, for $i=1,\cdots, M+1$. We assume without loss of generality that problem (\ref{73}) is feasible given the caching placement $\mathbf{X}$.\footnote{The feasibility of (\ref{73}) can be guaranteed in the alternating minimization method proposed in Section \ref{Alt}.} In the following Lemma $1$, we reveal an interesting structure of the feasible solutions of (\ref{73}) to simplify the problem.

\textbf{Definition 3:}  With a given $\mathbf{X}$, we refer to a block of consecutive tasks with $x_{i,j}=1$ for a specific program $j$ as a \textit{run}. Beside, we denote the index set of the first task of each run as $\mathcal{S}$, i.e., $\mathcal{S} = \left\{i | x_{i,j} > x_{i-1,j}, \forall i, j\right\}$. For instance, there are in total $3$ runs in Fig.~\ref{CacheExample}(b) (red boxes) and $\mathcal{S} = \left\{2,4,7\right\}$.

\textbf{Lemma 1:} With the given $\mathbf{X}$ in (\ref{73}), a necessary and sufficient condition for an offloading decision $\mathbf{a} \in \{0,1\}^M$ being a feasible solution of (\ref{73}) is that $a_{i-1} =1$, $\forall i \in \mathcal{S}$.

\emph{Proof:} Please see the detailed proof in Appendix C. $\hfill \blacksquare$

From Lemma $1$, we can equivalently replace all the constraints in (\ref{53}) with $a_{i-1} =1$, $\forall i \in \mathcal{S}$, which essentially removes $|\mathcal{S}|$ variables as well as all the $MN$ constraints in (\ref{53}). Then, we introduce auxiliary variables $b_i = a_{i-1}a_i$, $i=1,\cdots, M$, as in (\ref{76}), which transforms (\ref{73}) to the following ILP:
 \begin{equation}
 \label{81}
   \begin{aligned}
    & \underset{\mathbf{a,b}}{\text{minimize}} & & \mathsmaller \sum_{i=1}^{M}  \left(o_i^* + \lambda_i^* + \beta\tau_i^c + \beta \tau_{i+1}^d -l_i^*\right) a_i  \\
    & & &  - \mathsmaller \sum_{i=1}^{M} \left(o_i^* + \beta \tau_i^d\right) b_i  + \mathsmaller \sum_{i=1}^M l_i^*,\\
    & \text{subject to}  & &  b_i \leq  \frac{1}{2}\left(a_{i-1}+a_i\right), \ i=2,\cdots,M, \\
    & & &          a_{i}, b_i \in \left\{0,1\right\},\ i=1,\cdots,M, \\
    & & &          a_{i-1} =1,\ \forall i\in \mathcal{S}.
   \end{aligned}
\end{equation}
In general, the problem has $2M - |\mathcal{S}| $ binary variables. In the following, we study the properties of optimal offloading decisions to further reduce the complexity of solving the ILP in (\ref{73}).

\begin{figure*}
\normalsize
\setcounter{mytempeqncnt}{\value{equation}}
\setcounter{equation}{26}
\begin{equation}
\label{54}
\phi_{k,1}= \begin{cases}
o_1^* a_1 ,  &   k=1,\\
\sum_{i=e_{k} +1}^{u_k -1} \psi_i + \left[o_{u_k}^* \left(1-a_{u_k-1}\right)a_{u_k} + \beta \tau_{u_k}^d a_{u_k-1}\left(1-a_{u_k}\right)\right]   &   k=2,\cdots,K,\\
\end{cases}
\end{equation}

\begin{equation}
\label{55}
\phi_{k,0} = \left[ l_{u_k}^* \left(1-a_{u_k}\right) + \left(\lambda_{u_k}^*+ \beta \tau_{u_k}^c\right) a_{u_k} \right] + \mathsmaller\sum_{i=u_k +1}^{e_{k+1}} \psi_i,\ \ k=1,\cdots,K.
\end{equation}

\setcounter{equation}{\value{mytempeqncnt}}
\hrulefill
\vspace*{4pt}
\end{figure*}

Recall that the value of parameter $x_{i,\varphi_i}$ represents whether the program for computing the $i$th task is already in the service cache, where $i=1,\cdots,M$. Specifically, we refer to the $i$th task with $x_{i,\varphi_i}=1$ as a \emph{cached task}, and an \emph{uncached task} otherwise. To facilitate illustration, we set $x_{0,\varphi_0} = 1$ and $x_{M+1,\varphi_{M+1}}=0$ for the two virtual tasks without affecting both the objective and constraints of (\ref{73}). As an illustrative example in Fig.~\ref{CacheExample}(c), the caching state vector $\left[x_{0,\varphi_0},x_{1,\varphi_1},\cdots,x_{M+1,\varphi_{M+1}}\right]$ consists of alternating patterns of consecutive $0$'s and $1$'s. Here, we refer to a block of consecutive tasks with $x_{i,\varphi_i}=1$ as a \emph{cached segment}, and a block of consecutive tasks with $x_{i,\varphi_i}=0$ as an \emph{uncached segment}, such as the three cached segments and three uncached segments in Fig.~\ref{CacheExample}(c). In particular, these $M+2$ tasks always start with a cached segment and end with an uncached segment. Therefore, the numbers of cached and uncached segments are always the same and are denoted by $K\geq 1$. In the following, we separate our discussions according to the value of $K$.

\subsubsection{$K=1$} Note that $x_{1,\varphi_1}=0$ always holds because the service cache is assumed empty initially. Therefore, the first cached segment always has only one task (i.e., task $0$). $K=1$ indicates that $x_{i,\varphi_i} =0$ for $i=1,\cdots,M+1$. This indeed is the most difficult case in that we need to solve a general ILP by setting $\lambda_i^* = s_i^*$ for all $i=1,\cdots,M$ in (\ref{81}) without any improvement on computational complexity. In practice, however, this case rarely occurs when a proper initial caching placement is set.

\subsubsection{$K>1$} In this case, there exists some cached task $i$ for $1< i \leq M$. We denote $e_k$ and $u_k$ as the indices of the uncached tasks preceding and following the $k$th cached segment, respectively, while $e_1$ is not defined. For instance, $u_1=1$, $\{e_2,u_2\} = \{3,6\}$ and $\{e_3,u_3\} = \{6,11\}$ in Fig.~\ref{CacheExample}(c). Notice that $u_{k} = e_{k+1}$ may occur when there is only one task in an uncached segment, such as $u_2=e_3=6$. For simplicity of illustration, we denote
\begin{equation*}
\label{56}
\begin{aligned}
\psi_i =& o_i^* \left(1-a_{i-1}\right)a_i  + l_i^* \left(1-a_i\right) \\
&+ \left(\lambda_i^*+ \beta \tau_i^c\right) a_i   + \beta \tau_i^d a_{i-1}\left(1-a_i\right),
\end{aligned}
\end{equation*}
such that the objective of (\ref{73}) is expressed as $\sum_{i=1}^{M+1} \psi_i$. Alternatively, the objective of (\ref{73}) can be decomposed based on $\{e_k,u_k\}$'s, such that problem (\ref{73}) can be recast as following
 \begin{subequations}
 \label{74}
   \begin{align}
    \ \ \ & \underset{\mathbf{a}\in\{0,1\}^M}{\text{minimize}}  & & \mathsmaller\sum_{k=1}^K \left(\phi_{k,1} + \phi_{k,0}\right)\\
    & \text{subject to}  & &  a_{i-1} =1,\ \forall i\in \mathcal{S},
   \end{align}
\end{subequations}
where $\phi_{k,1}$ and $\phi_{k,0}$ are expressed in (\ref{54}) and (\ref{55}) respectively at the top of this page. Intuitively, $\phi_{k,1}$ and $\phi_{k,0}$ correspond to the TEC induced by the $k$th cached and uncached segments, respectively. Besides, the sets of optimizing variables in $\phi_{k,1}$ and $\phi_{k,0}$ are
\setcounter{equation}{28}
\begin{equation*}
\label{57}
\mathcal{A}_{k,1} = \begin{cases}
a_1 ,  &   k=1,\\
\left\{a_i | i = e_{k}, e_{k}+1, \cdots, u_k \right\},   &   k=2,\cdots,K,\\
\end{cases}
\end{equation*}
and $\mathcal{A}_{k,0} = \left\{a_i | i = u_k, u_k+1, \cdots, e_{k+1} \right\}, \ k=1,\cdots,K.$

\begin{figure}
\centering
  \begin{center}
    \includegraphics[width=0.48\textwidth]{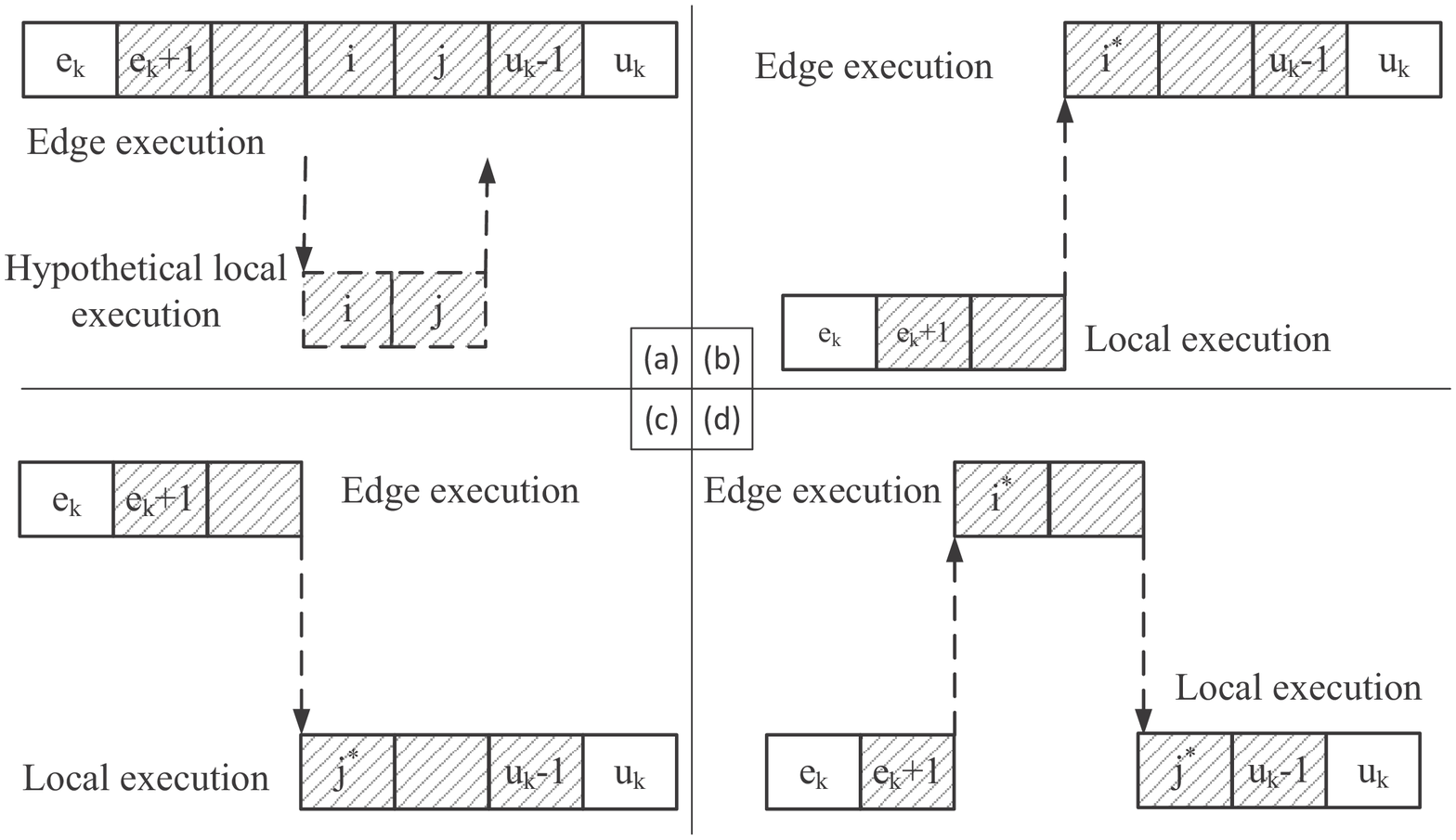}
  \end{center}
  \caption{Optimal offloading decision given the values of $a_{e_{k}}$ and $a_{u_k}$.  }
  \label{CachedTasks}
\end{figure}

A closer observation of $\mathcal{A}_{k,1}$ and $\mathcal{A}_{k,0}$ shows that once the values of $\{a_{e_k},a_{u_k}\}$'s are fixed, for $i=1,\cdots,K$, $\phi_{k,1}$'s and $\phi_{k,0}$'s can be separately optimized with disjoint sets of variables. In the following, we first discuss the optimal offloading decisions of the cached tasks that minimize $\phi_{k,1}$'s. Without loss of generality, we focus on the $k$th cached segment, supposing that $\{a_{e_{k}}, a_{u_k}\}$ are given. Depending on the values of $\{a_{e_{k}}, a_{u_k}\}$, there are four cases, as illustrated in Fig.~\ref{CachedTasks}.
\begin{enumerate}
  \item $a_{e_{k}} =a_{u_k}=1$, as shown in Fig.~\ref{CachedTasks}(a). In this case, the optimal offloading solution is $a_i=1$, for $i=e_{k}+1,\cdots,u_k-1$. That is, all the cached tasks are executed at the edge server. Due to the page limit, we only provide a sketch of proof here, by contradiction. Suppose that tasks $i$ to $j$ are computed locally instead of at the edge in Fig.~\ref{CachedTasks}(a). This will incur not only additional time and energy for downloading (uploading) the input (output) of the $i$th ($j$th) task, but also additional time and energy for local computation, because $f_0>f_{max}$ and the energy consumption on edge computation is neglected.
  \item $a_{e_{k}}=0$ and $a_{u_k}=1$, as shown in Fig.~\ref{CachedTasks}(b). For the optimal offloading decision, there must exist an optimal task $i^* \in \left\{e_{k}+1,\cdots,u_k-1\right\}$, such that for each $i = e_{k}+1,\cdots,u_k-1$, we have $a^*_i =0$ if $i<i^*$ and $a^*_i =1$ if $i \geq i^*$. This indicates that the computation is offloaded to the edge exactly once within the segment. The proof follows that in the first case and is omitted for brevity. In particular, $i^*$ can be found via a simple linear search.
  \item $a_{e_k}=1$ and $a_{u_k}=0$, as shown in Fig.~\ref{CachedTasks}(c). For the optimal offloading decision, there must exist an optimal task $j^* \in \left\{e_{k}+1,\cdots,u_k-1\right\}$, such that for each $i = e_{k}+1,\cdots,u_k-1$, we have $a^*_i = 1$ if $i < j^*$ and $a^*_i = 0$ if $i \geq j^*$. The proof also follows the idea in the first case. This indicates that the computation result is downloaded to the MU exactly once within the segment. In particular, $j^*$ can be found using a linear search.
  \item $a_{e_{k}} =a_{u_k}= 0$, as shown in Fig.~\ref{CachedTasks}(d), implying that the computations start and end both at the MU. This corresponds to the case in \cite{2018:Yan}, which shows that the optimal computation offloading strategy satisfies a ``one-climb" policy where the tasks are either offloaded to the edge server for exactly once, or all executed locally at the MU. There must exist $i^*\leq j^*$, such that the optimal solution of $a_i$, $i = e_{k}+1,\cdots,u_k-1$, is
\begin{equation}
a_i^* = \begin{cases}
0 ,  &   i < i^* \text{ or }  i \geq j^*, \\
1 ,  &   i^* \leq i < j^*.
\end{cases}
\end{equation}
The optimal $\{i^*,j^*\}$ can be efficiently obtained through a two-dimensional search.
\end{enumerate}

From the above discussion, the optimal value $\phi_{k,1}$ under the above four cases can be efficiently obtained. Let us denote the optimal values by $v_k^{(1)}$, $v_k^{(2)}$, $v_k^{(3)}$, and $v_k^{(4)}$ for the four cases, respectively. Moreover, the calculations of $\left\{v_k^{(1)}, v_k^{(2)}, v_k^{(3)}, v_k^{(4)}\right\}$'s can be performed in parallel for different segments. This way, $\phi_{k,1}$ can be expressed as
\begin{equation}
\label{66}
\begin{aligned}
\phi_{k,1} &= v_k^{(1)} a_{e_k} a_{u_k} + v_k^{(2)} \left(1- a_{e_k}\right)a_{u_k}  \\
&+ v_k^{(3)} a_{e_k}\left(1-a_{u_k}\right) + v_k^{(4)} \left(1- a_{e_k}\right)\left(1-a_{u_k}\right).
\end{aligned}
\end{equation}
By substituting (\ref{66}) into (\ref{74}), we eliminate all the offloading decision variables corresponding to the cached tasks, and leaving only the variables for the uncached tasks, i.e., $\{a_i| x_{i,\varphi_i} =0, i=1,\cdots, M\}$. In the following, we transform (\ref{74}) into an equivalent ILP problem.

\subsection{Equivalent ILP Formulation}\label{OffOpt}
\begin{figure}
\centering
  \begin{center}
    \includegraphics[width=0.45\textwidth]{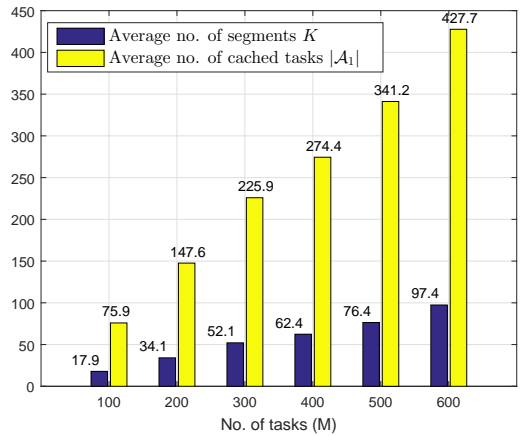}
  \end{center}
  \caption{The average number of segments $K$ and cached tasks $|\mathcal{A}_1|$ when solving (\ref{75}) during the alternating minimization under different number of tasks $M$. The simulation parameters are in Table I.}
  \label{num_seg}
\end{figure}

The basic idea is similar to that for (P3) in Section~\ref{Joint}, where the new challenge is in the multiplicative terms in (\ref{66}). By denoting $\hat{a}_i \triangleq 1 - a_i$, where $\hat{a}_i \in\{0,1\}$, we rewrite (\ref{66}) as
\begin{equation}
\begin{aligned}
\phi_{k,1} &= v_k^{(1)} a_{e_k} \left(1 - \hat{a}_{u_k}\right) + v_k^{(2)} \left(1- a_{e_k}\right)a_{u_k} \\
&+ v_k^{(3)} a_{e_k}\left(1-a_{u_k}\right) + v_k^{(4)} \left(1- a_{e_k}\right)\hat{a}_{u_k}.
\end{aligned}
\end{equation}
We further define $q_k \triangleq a_{e_k} a_{u_k}$ and $\hat{q}_k \triangleq a_{e_k} \hat{a}_{u_k}$, and express the above equation as
\begin{equation}
\label{67}
\begin{aligned}
\omega_{k,1} &\triangleq \left(v_k^{(1)}+ v_k^{(3)}\right) a_{e_{k}} + v_k^{(2)} a_{u_k} + v_k^{(4)} \hat{a}_{u_k}\\
&- \left(v_k^{(1)} + v_k^{(4)}\right)\hat{q}_k - \left(v_k^{(2)} + v_k^{(3)}\right)q_k.
\end{aligned}
\end{equation}
By substituting (\ref{67}) into (\ref{74}) and introducing auxiliary variables $b_i = a_{i-1}a_{i}$, we have
 \begin{subequations}
 \label{75}
   \begin{align}
    \ \ \ & \underset{\mathbf{a},\hat{\mathbf{a}},\mathbf{b},\mathbf{q},\hat{\mathbf{q}}}{\text{minimize}} & & \mathsmaller \sum_{k=1}^K \left(\omega_{k,1} + \phi_{k,0}\right)\\
    & \text{subject to}  & &  a_{i-1} =1,\ \forall i\in \mathcal{S},\\
    & & &  b_i \leq  \frac{1}{2}\left(a_{i-1}+a_i\right), \label{45}\\
    & & &  \ \ \forall i\in \mathcal{A}_{k,0}\setminus u_k, k=1,\cdots,K, \\
    & & &  q_k \leq  \frac{1}{2}\left(a_{e_{k}} + a_{u_k} \right), k = 2,\cdots,K,\\
    & & &  \hat{q}_k \leq  \frac{1}{2}\left(a_{e_k} + \hat{a}_{u_k} \right), k = 2,\cdots,K,  \label{46}\\
    & & &  \hat{a}_{u_k} + a_{u_k} = 1, \ k=2, \cdots, K, \\
    & & &  a_i, \hat{a}_i, b_i, q_k, \hat{q}_k \in \{0,1\}, \ \forall i,k.
   \end{align}
\end{subequations}
We see that the inequalities (\ref{45}) to (\ref{46}) are equivalent to $b_i = a_{i-1}a_{i}$, $q_k = a_{e_k} a_{u_k}$, and $\hat{q}_k = a_{e_k} \hat{a}_{u_k}$, respectively, because the objective decreases with $\{b_i, q_k, \hat{q}_k\}$'s. Similar to (\ref{81}), the problem above is also a pure 0-1 integer optimization problem. Compared to (\ref{81}), it reduces $2|\mathcal{A}_1|$ variables that correspond to the cached tasks, where $\mathcal{A}_1 = \left\{i| x_{i,\varphi_i} =1, i=1,\cdots, M\right\}$, while introducing additional $3(K-1)$ auxiliary variables. In general, the above formulation can effectively reduce the computational complexity because $|\mathcal{A}_1|$ is often much larger than $K$ in practice. To see this, we plot in Fig.~\ref{num_seg} the average number of cached tasks ($|\mathcal{A}_1|$) and segments ($K$) when solving (\ref{75}) during the execution of the alternating minimization (to be introduced in the next subsection), where the former is more than $4$ times larger than the latter for all $M$. The complexity of the alternating minimization will be evaluated numerically in Section VI.B.

\subsection{Alternating Minimization}\label{Alt}
Sections \ref{CacheOpt} and \ref{decomp} show that we can compute the optimal caching placement $\mathbf{X}^*$ with low complexity when the offloading decision $\mathbf{a}$ is given, and vice versa. This leads to an alternating minimization scheme that optimize the two set of variables $\mathbf{X}$ and $\mathbf{a}$ alternately. Starting from an initial $\mathbf{a}^{(0)}$, we iteratively compute the optimal $\mathbf{X}^{(i)}$ given $\mathbf{a}^{(i-1)}$ (by solving (P4)), and the optimal $\mathbf{a}^{(i)}$ given $\mathbf{X}^{(i)}$ (by solving problem (\ref{73})) for $i=1,2,\cdots$, until the improvement on the objective function of (P3) becomes marginal. Because the objective of (P3) is bounded below and non-increasing as the iterations proceed, the alternating minimization method is asymptotically convergent. The detailed algorithm description is omitted for brevity.

\section{Simulation Results}
In this section, we evaluate the performance of the proposed algorithms through numerical simulations. All the computations are solved in MATLAB on a computer with an Intel Core i7-4790 3.60-GHz CPU and 16 GB of memory. Besides, we use Gurobi optimization tools to solve the ILP problems \cite{2019:Gurobi}. In all simulations, we assume that the average channel gain $\bar{h}_i$ follows a path-loss model $\bar{h}_i = A_d\left(\frac{3\times 10^8}{4\pi f_c d_M}\right)^{d_e}$, $i=1,\cdots,M$, where $A_d$ denotes the antenna gain, $f_c$ denotes the carrier frequency, $d_e$ denotes the path loss exponent, and $d_M$ denotes the distance between the MU and the edge server. The time-varying fading channel $h_i$ follows an i.i.d. Rician distribution with LOS link gain equal to $0.2 \bar{h}_i$. Unless otherwise stated, the parameters used in the simulations are listed in Table \ref{tab:parameter}, which correspond to a typical outdoor MEC system. For simplicity of illustration, we assume that $c_j$'s are equal for all the programs, such that the caching capacity $C$ is normalized to indicate the number of programs that can cache.

\begin{table}
\caption{Simulation Parameters}
\footnotesize
\begin{center}
\begin{tabular}{|l|| l| }
\hline
  $B = \bar{B}= 10^6$ Hz                      & $f_{max} = 0.5$ GHz    \\ \hline
  $\sigma^2 = \bar{\sigma}^2=10^{-10}$ Watt   &   $\kappa = 10^{-26}$  \\ \hline
  $P_0 = 1$ Watt               &   $\beta = 0.1$                       \\ \hline
  $f_0 = 10$ GHz               &   $M = 400$                           \\ \hline
  $P_{max} = 0.1$ Watt         &   $N =6$                              \\ \hline
  $O_i \in [2, 5]$ Mb                 & $d_M =30$ meters               \\ \hline
 $L_i \in [50,200]\times 10^6$ Cycles & $d_e = 2.6$                    \\ \hline
 $s_j \in [0.5,1.5]$ Mb              & $A_d =4.11$                     \\ \hline
  $D_j = 3$ seconds, $\forall j$      & $f_c =915$ MHz                 \\ \hline
  Normalized $C =3$                   &  $\alpha=3$                    \\ \hline
\end{tabular}
\end{center}
\label{tab:parameter}
\end{table}

Unless otherwise stated, all results in the simulations are the average performance of $50$ independent simulations. In each simulation, we first randomly generate $M$ tasks that belong to $N=6$ types of programs, where the types of the sequential tasks follow a Markov chain with a random initial state. Specifically, the Markov transition probability $P_{i,j} \triangleq \text{Pr}\left(\varphi_{k+1} =j|\varphi_k = i\right) = 0.4$ if $i=j$, and $P_{i,j} = 0.12$ if $i\neq j$, $\forall i,j,k$, where $\varphi_k$ denotes the program type of the $k$th task. Then, the parameters of each task ($O_i$ and $L_i$) and each type of program ($s_j$) are uniformly generated from the ranges specified in Table \ref{tab:parameter} for $i = 1,\cdots,M$ and $j=1,\cdots,N$.

In the following, we evaluate the performance of the proposed optimal joint optimization (in Section \ref{Joint}) and alternating minimization (in Section \ref{Alt}) methods. Specifically, we initialize $a_i =1$ for all $i$ in the alternating minimization. Besides, we also consider the following benchmark methods for performance comparison:
\begin{itemize}
  \item Popular-cache: we first neglect the offloading decision of the MU and cache the most popular programs that are executed most frequently throughout the time. Then, we optimize the offloading decision given the caching placement using the method in Section \ref{OffOpt}. This is similar to the \textit{top-R caching} method in \cite{2018:He} and \textit{Pop-aware caching} method in \cite{2018:Shukla}.
  \item Cache-oblivious offloading: we first neglect the edge caching placement and optimize the offloading decisions by assuming that the service programs for processing all the tasks are available at the edge server. Then, we optimize the caching placement based on the obtained offloading decision using the method in Section \ref{CacheOpt}.
\end{itemize}

\begin{figure}
\centering
  \begin{center}
    \includegraphics[width=0.45\textwidth]{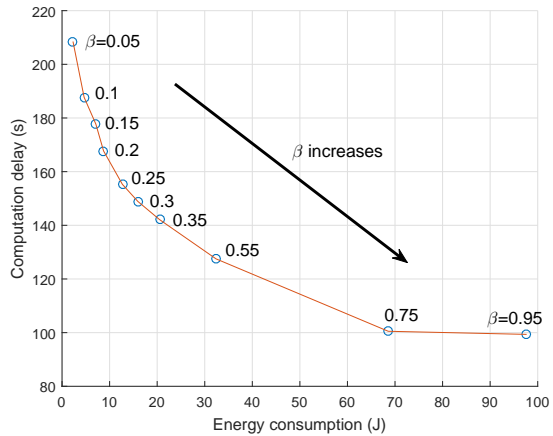}
  \end{center}
  \caption{The optimal energy-delay tradeoff of the joint optimization under different values of $\beta$.}
  \label{tradeoff}
\end{figure}

\subsection{TEC Performance Evaluation}
We first examine the optimal performance tradeoff between the two competing objectives in (P1), i.e., computation delay $T$ and energy consumption $E$ of the MU, by varying the weighting parameter $\beta$. As excepted, $T$ decreases with $\beta$ while $E$ increases. In particular, $T$ first quickly decreases with $\beta$ and gradually becomes a constant when $\beta \geq 0.75$. The curve can be used to set proper value of $\beta$, e.g., setting $\beta =0.1$ when requiring the total energy consumption to be lower than $5$ J. Without loss of generality, we set $\beta =0.1$ in the following simulations.

\begin{figure*}
\centering
  \begin{center}
    \includegraphics[width=1\textwidth]{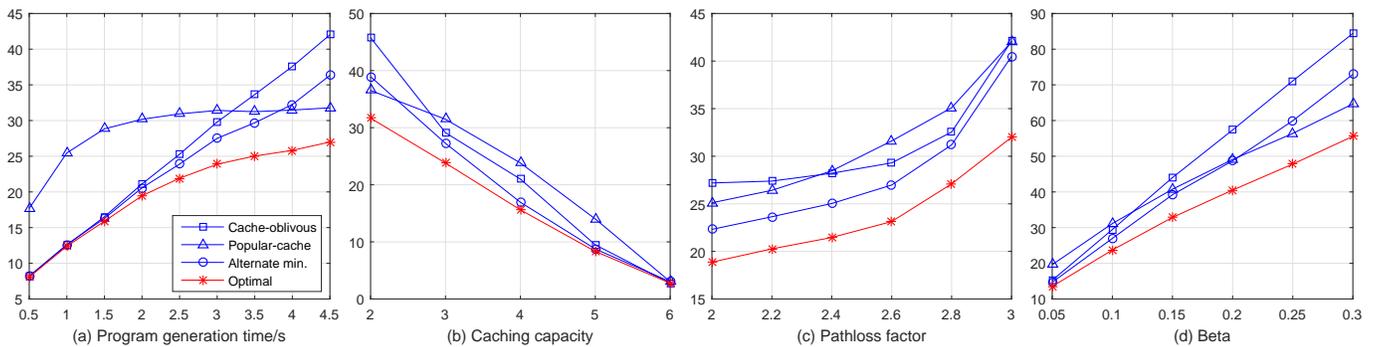}
  \end{center}
  \caption{TEC performance comparisons of different methods.}
  \label{TEC}
\end{figure*}

\begin{figure*}
\centering
  \begin{center}
    \includegraphics[width=1\textwidth]{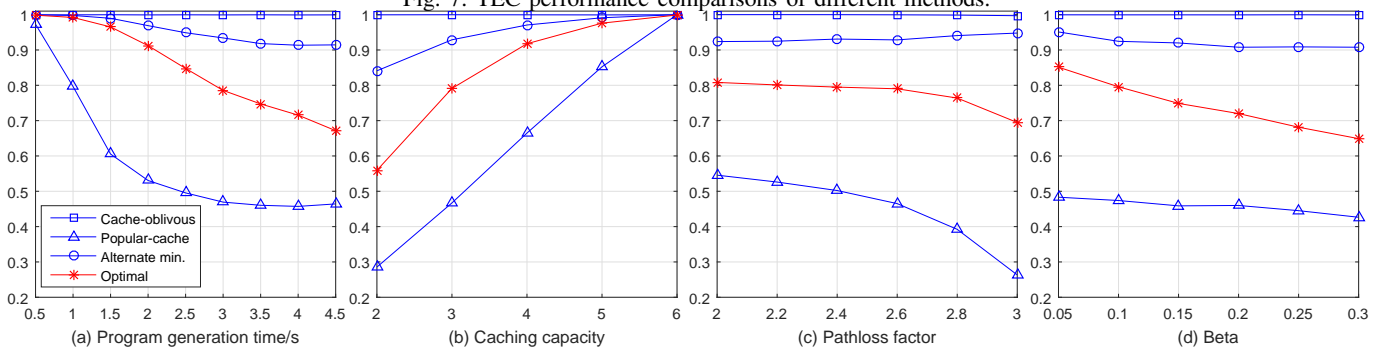}
  \end{center}
  \caption{Ratio of offloaded tasks when different methods are applied.}
  \label{Ratio}
\end{figure*}

We then evaluate the TEC performance under different system setups. In Fig.~\ref{TEC}(a), we vary the program generation time $D_j$ from $0.5$ to $4.5$ seconds, which naturally results in an increase of TEC for all the methods. The Cache-oblivious method performs closely to the optimal scheme when $D_j$ is small, e.g., $D_j \leq 2$, but its performance degrades as $D_j$ further increases. To examine the underlying cause, we plot in Fig.~\ref{Ratio}(a) the ratio of offloaded tasks. As expected, the offloading ratios of all methods decrease with the program generation time. Meanwhile, we notice that the Cache-oblivious method offloads almost all the tasks for edge execution under different $D_j$, which is consistent with the one-climb offloading policy in \cite{2018:Yan}. As a result, the conserved computation time and energy from edge computation is gradually exceeded by the overhead due to frequent offloading/initialization of new programs when $D_j$ increases. The alternating minimization method has a similar trend as the Cache-oblivious method, because it is largely affected by the initial offloading solution where all the tasks are offloaded to the edge. Besides, the TEC performance of the Popular-cache method gradually converges as $D_j$ increases. This is because when $D_j$ is large, a task tends to be offloaded for edge execution only if its required program is already cached. As popular-cache method has a fixed caching placement throughout the time ($C=3$ out of the $N=6$ programs are cached), its offloading ratio converges to $0.5$ when $D_j$ is large and is verified in Fig.~\ref{Ratio}(a). This also leads to a convergent TEC performance.

In Fig.~\ref{TEC}(b) and Fig.~\ref{Ratio}(b), we vary the normalized caching capacity from $2$ to $6$. Because a larger caching capacity translates to more savings in program offloading and generation, the TEC decreases and the task offloading ratio increases for all the schemes considered. Specifically, when $C=6$, i.e., all the programs can be stored in the cache, the Cache-oblivious scheme approaches the optimal scheme. In Fig.~\ref{TEC}(c) and Fig.~\ref{Ratio}(c), we vary the path-loss factor $d_e$ from $2$ to $3$, which leads to a drastic decrease of wireless channel gains. As expected, the weaker channels suffer lower offloading ratio and higher TEC because the higher cost of transmitting the task and program data discourages task offloading. The proposed joint optimization has significant performance gain over all the other schemes considered, especially when $d_e$ is large. Specifically, it reduces the TEC by more than $25\%$ compared with all the other schemes when $d_e = 3$.

At last, in Fig.~\ref{TEC}(d) and Fig.~\ref{Ratio}(d), we vary the weighting parameter $\beta$ from $0.05$ to $0.3$, where a smaller (larger) $\beta$ indicates stronger emphasis on minimizing the energy consumption (delay). We notice that the TEC increases with $\beta$ because the delay cost dominates the energy consumption (e.g., one order of amplitude larger in Fig.~\ref{tradeoff} when $\beta\leq 0.3$). Meanwhile, the high program generation delay discourages task offloading, leading to a decreased offloading ratio when $\beta$ increases. As a result, the Cache-oblivious scheme performs the worst when $\beta =0.3$ because of the high delay cost on program generation at the edge server.

Overall, the optimal joint optimization scheme has evident TEC performance advantage over the others schemes. The Cache-oblivious scheme performs well only when the program generation time is short or under less stringent delay requirement. The Popular-cache scheme performs poorly in most cases due to its negligence to the task offloading decisions. This is in contrast to the traditional content caching schemes, where caching popular contents (e.g., large and most frequently accessed files) usually performs well. The alternating minimization has relatively good performance in most scenarios. However, more tasks are offloaded than actually required in the optimal solution.

\begin{figure}
\centering
  \begin{center}
    \includegraphics[width=0.47\textwidth]{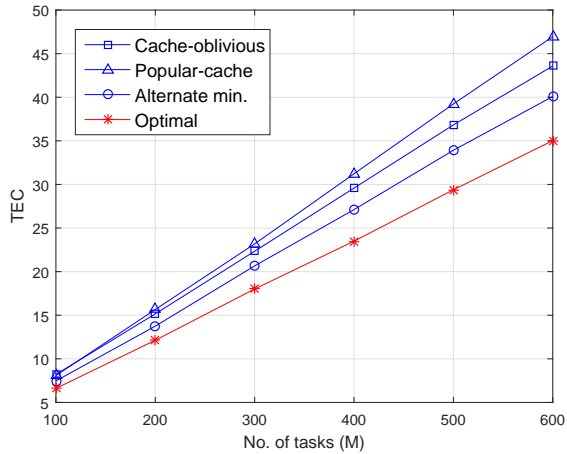}
  \end{center}
  \caption{TEC performance comparison when the number of tasks varies.}
  \label{TEC_num}
\end{figure}

We also compare in Fig.~\ref{TEC_num} the TEC Performance when the number of tasks $M$ varies. We observe that the TEC increases linearly with $M$ for all the schemes, while the optimal scheme and the alternating minimization significantly outperforms the others. In particular, the optimal scheme achieves on average $13.5\%$ lower TEC than the alternating minimization method. In the following, we evaluate the computational complexity of the two best-performing schemes, i.e., the joint optimization and the alternating minimization methods.

\subsection{Complexity Evaluation}
\begin{figure}
\centering
  \begin{center}
    \includegraphics[width=0.47\textwidth]{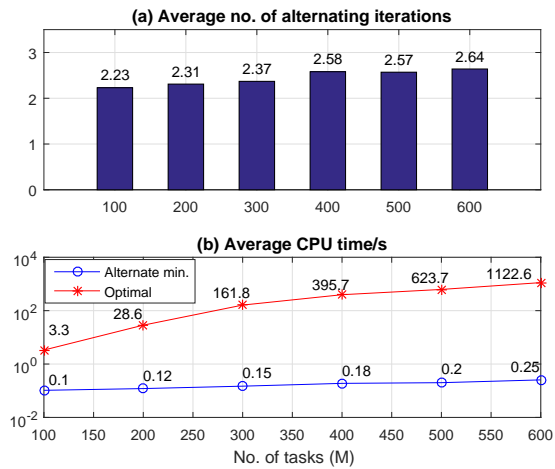}
  \end{center}
  \caption{Comparison of CPU time when the number of tasks varies.}
  \label{complexity}
\end{figure}

When the number of tasks $M$ varies from $100$ to $600$, we plot in Fig.~\ref{complexity} the average number of iterations used by alternating minimization and the average CPU time comparison of the joint optimization and the alternating minimization methods. The result is an average of $100$ independent simulations. We see in Fig.~\ref{complexity}(a) that the average number of iterations of the alternating minimization method does not vary significantly and is below $3$ for all $M$. This is also one important reason behind the slow increase of CPU time of the alternating minimization method in Fig.~\ref{complexity}(b), where the CPU time increases slightly from $0.1$ to $0.25$ second when $M$ increases by $6$ times. In vivid contrast, the CPU time of the joint optimization method increases by more than $340$ times from $3.3$ seconds to around $19$ minutes. The quick increase of CPU time may result in an unaffordable delay in practice when $M$ is large. In practice, the alternating minimization method provides a reduced-complexity alternative.

\section{Conclusions and Future Works}
In this paper, we have considered a cache-assisted single-user MEC system, where the server can selectively cache the previously generated programs for future reuse. To minimize the computation delay and energy consumption of the MU, we studied the joint optimization of service caching placement, computation offloading decisions, and system resource allocation. We first transformed the complicated MINLP problem to a pure 0-1 ILP problem by separately deriving the closed-form expressions of the optimal resource allocation. Then, we proposed reduced-complexity algorithms to obtain the optimal caching placement by fixing the offloading decision, and vice versa. We further devised an alternating minimization to update the caching placement and offloading decision alternately. Extensive simulations show that the joint optimization achieves substantial resource savings of the MU compared to other representative benchmark methods considered. In particular, the sub-optimal alternating minimization method achieves a good balance of system performance and computational complexity.

Finally, we conclude the paper with some interesting future working directions of service-cache assisted MEC. First, it is interesting to consider dynamic resource allocation (on computing, storage, and communication resource) in a general multi-user MEC system to improve utilization efficiency. In particular, the cached service programs from different MUs may be shared to reduce the program uploading cost. This also raises many new technical challenges, such as inter-user interference in task offloading and privacy issues in service sharing. Secondly, it is also promising to extend the single-server setup to a multi-server one. For instance, we may consider a two-tier MEC network where some large-size computation tasks are forwarded by local micro-BS edge server to more powerful macro-BS server. Meanwhile, we can balance the computation workloads at different edge servers by allowing multiple edge servers to provide computing service collaboratively. Thirdly, we assume in this paper a sequential task execution model. Moving forward, it is of high practical value to study cache-assisted computations under a general task execution model, e.g., a tree or mesh model. Fourthly, it is worth studying a scenario where the edge platform can download some publicly accessible programs from the core network, such that it may pre-cache the service data beforehand to further reduce computation delay. Fifthly, the cache update frequency of this work is relatively high due to the dynamic task arrivals, e.g., on average once every $22.4$ seconds of the optimal scheme in the simulation. When cache switching cost is considered, additional penalty terms need to be included in the objective to reduce the update frequency. At last, we studied in this paper an offline optimization problem that assumes the future system parameters, such as the wireless channel gains and task data size, are known beforehand. In practice, they may be revealed only upon the task executions, thus an online design is needed. Depending on the knowledge of future information, there are numerous methods to design an online scheme. For instance, when the channel and task arrival distributions are known, we can apply dynamic programming technique to minimize the expected cost. Otherwise, when they are unknown, we may apply reinforcement learning technique \cite{2019:Huang} to directly learn the optimal mapping between the caching/channel state to the offloading and caching actions.

\begin{appendices}

\section{Proof of Proposition $1$}
\emph{Proof:} We denote the objective of the problem as $L(\tau^u_i)$, which is a strictly convex function within the feasible set $\tau^u_i \geq \frac{O_{i-1}}{R_i^{max}} \geq 0$. Accordingly, the minimum is achieved at either the boundary point $\frac{O_{i-1}}{R_i^{max}}$ or the point $v_1$ that satisfies $L'(v_1)=0$, depending on the value of $v_1$. To obtain $v_1$, we take the derivative of $L(\tau^u_i)$ and set it equal to zero, i.e.,
\begin{equation}
\label{19}
\begin{aligned}
&L'(\tau^u_i) \\
=& \beta + \frac{(1-\beta)\sigma^2}{h_i} \left(2^{\frac{O_{i-1}}{B \tau^u_i}}-1 - \ln 2 \cdot 2^{\frac{O_{i-1}}{B \tau^u_i}} \cdot \frac{O_{i-1}}{B \tau^u_i} \right)\\
=& \frac{(1-\beta)\sigma^2 e}{h_i} \bigg[ e^{-1}\left(\frac{\beta h_i}{(1-\beta)\sigma^2} -1\right)  \\
&- e^{\ln 2 \frac{O_{i-1}}{B \tau^u_i}- 1}\left(\ln 2 \cdot\frac{O_{i-1}}{B \tau^u_i} - 1\right)\bigg]=0,\\
\Rightarrow& e^{\ln 2 \frac{O_{i-1}}{B \tau^u_i}- 1}\left(\ln 2 \cdot\frac{O_{i-1}}{B \tau^u_i} - 1\right) = e^{-1}\left(\frac{\beta h_i}{(1-\beta)\sigma^2} -1\right).
\end{aligned}
\end{equation}
Because $e^{-1}\left(\frac{\beta h_i}{(1-\beta)\sigma^2} -1\right)\geq -1$, the above equality is equivalent to
\begin{equation}
\ln 2 \cdot \frac{O_{i-1}}{B \tau^u_i}- 1 = \mathcal{W}\left(e^{-1}\left[\frac{\beta h_i}{(1-\beta)\sigma^2} -1\right]\right),
\end{equation}
where $\mathcal{W}(x)$ denotes the Lambert-W function. Therefore, we have $v_1 = \frac{\ln 2 \cdot O_{i-1}}{B\cdot \left[  \mathcal{W}\left(e^{-1}\left[\frac{\beta h_i}{(1-\beta)\sigma^2} -1\right]\right)+ 1\right]}$.

If $v_1 < \frac{O_{i-1}}{R_i^{max}}$, or equivalently $L'(\tau^u_i)=0$ is not achievable within the feasible set, we can infer that the optimal solution is obtained the boundary $\left(\tau_i^u\right)^* = \frac{O_{i-1}}{R_i^{max}}$. Because $L(\tau^u_i)$ is convex, $L'(\tau^u_i)$ is an increasing function. Given $L'(v_1)=0$, the condition $v_1 < \frac{O_{i-1}}{R_i^{max}}$ is equivalent to $L'\left(\frac{O_{i-1}}{R_i^{max}}\right)>0$. By substituting $\tau^u_i = \frac{O_{i-1}}{R_i^{max}}$ into (\ref{19}), we have $v_1 < \frac{O_{i-1}}{R_i^{max}}$ when
\begin{equation}
\label{21}
\begin{aligned}
&\beta + (1-\beta)P_{max} \left[ 1 - \ln\left(1+ q_i\right)\left(\frac{1}{q_i}+ 1\right) \right]  > 0\\
\Rightarrow& \ln\left(1+q_i\right)  \leq \left(1+ \frac{\beta}{(1-\beta)P_{max}}\right)\left(1- \frac{1}{1+q_i}\right)\\ \Rightarrow& \ln\left(\frac{1}{1+q_i}\right) \geq -A + \frac{A}{1+q_i},
\end{aligned}
\end{equation}
where $q_i \triangleq \frac{h_i P_{max}}{\sigma^2}$ and $A \triangleq 1+ \frac{\beta}{(1-\beta)P_{max}}$. By taking a natural exponential operation at both sides of (\ref{21}), we have
\begin{equation*}
\begin{aligned}
&\exp \left(-\frac{A}{1+q_i}\right) \left(\frac{1}{1+q_i}\right) \geq \exp\left(-A\right) \\
&\Rightarrow \exp \left(-\frac{A}{1+q_i}\right) \left(-\frac{A}{1+q_i}\right) \leq  - A\exp\left(-A\right).
\end{aligned}
\end{equation*}
Evidently, the RHS of the above inequality satisfies $ e^{-1} \leq - A\exp\left(-A\right) \leq 0$. Then, the above inequality can be equivalently expressed as
\begin{equation}
\label{23}
-\frac{A}{1+q_i} \leq \mathcal{W}\left( - A\exp\left(-A\right) \right).
\end{equation}
The equivalence holds because $\mathcal{W}(x)$ is an increasing function when $x\geq -1/e$. After some simple manipulation, we obtain from (\ref{23}) that the optimal solution $\left(\tau_i^u\right)^* = \frac{O_{i-1}}{R_i^{max}}$ when
\begin{equation}
\label{24}
h_i \leq \frac{\sigma^2}{P_{max}}\left(\frac{A}{-\mathcal{W}\left( - A\exp\left(-A\right) \right)}-1\right).
\end{equation}
Otherwise, if (\ref{24}) does not hold, we conclude that $v_1 \geq \frac{O_{i-1}}{R_i^{max}}$ and $L'(\tau^u_i)=0$ is achievable such that the optimal solution is $\left(\tau_i^u\right)^*=v_1$. This proves Proposition 1. $\hfill \blacksquare$

\section{Proof of Proposition $2$}
\emph{Proof:} Consider a tagged task $k \in \mathcal{A}$ and $\varphi_k =j$ (i.e, $a_k = u_{k,j} = 1$). We examine the potential change of feasibility and objective value of (P4) after setting $x_{i,j} = \hat{x}_{k,j}$, $\forall i \in \mathcal{V}_k$, in the solution of $\mathbf{X} = \mathbf{\hat{X}}$. We assume without loss of generality that $|\mathcal{V}_k|>1$. Because the value of $x_{k,j}$ remains unchanged, we only focus on task $i\in \mathcal{V}_k \setminus k$. By definition, for a task $i\in \mathcal{V}_k \setminus k$, either $a_i=0$ or $u_{i,j}=0$ must hold (i.e., $a_i u_{i,j}=0$), because otherwise we would have $\nu_{i}^j =i$, which contradicts with our assumption that $i \in \mathcal{V}_k \setminus k$.

We first examine the impact to the feasibility of (P4). If $\hat{x}_{k,j}=1$, the corresponding constraints in (\ref{101}) reduces to
\begin{equation}
\begin{cases}
x_{k-1,j} \geq 1,  &   \\
x_{i-1,j} \geq x_{i,j}, & i = k- |\mathcal{V}_k| +1,\cdots, k-1.
\end{cases}
\end{equation}
Using backward induction from $i=k-1$ to $i=k - |\mathcal{V}_k| +1$, we can infer that $x_{i,j} =1$ must hold $\forall i \in \mathcal{V}_k\setminus k$ to satisfy the above inequalities. On the other hand, if $\hat{x}_{k,j}= 0$, by setting $x_{i,j} =0$, $\forall i \in \mathcal{V}_k\setminus k$, we see that the corresponding constraints in (\ref{101}) reduces to $x_{i,j} \geq 0, \forall i \in \mathcal{V}_k$, which are automatically satisfied. Meanwhile, all the constraints in (\ref{100}) still hold because the LHS of (\ref{100}) is non-increasing when setting $x_{i,j} =0$, $\forall i \in \mathcal{V}_k \setminus k$. From the above discussion, (P4) is still feasible after setting $x_{i,j} = \hat{x}_{k,j}$, $\forall i \in \mathcal{V}_k \setminus k$.

Because the value of $x_{i,j}$, $\forall i \in \mathcal{V}_k\setminus k$, does not affect the objective value of (P4), the objective remains unchanged after setting $x_{i,j} = \hat{x}_{k,j}$, $\forall i \in \mathcal{V}_k \setminus k$. Therefore, after repeating the above substitutions for all the $k \in \mathcal{A}$, (P4) is still feasible and the objective remains unchanged, which proves Proposition 2. $\hfill \blacksquare$

\section{Proof of Lemma $1$}
\emph{Proof:} We first prove the necessary condition. That is, if $\mathbf{a} \in \{0,1\}^M$ is a feasible solution of (\ref{73}), then $a_{i-1} =1$, $\forall i \in \mathcal{S}$ must hold. Suppose that $\mathbf{a}$ is a feasible solution, with the given $\mathbf{X}$, the following constraints in (\ref{53}) are satisfied
\begin{equation}
      \label{5}
      u_{i-1,j} a_{i-1}  \geq x_{i,j} - x_{i-1,j}, \ \forall i \in \mathcal{S},\  j=1,\cdots,N.
\end{equation}
By the definition of $\mathcal{S} = \left\{i | x_{i,j} > x_{i-1,j}, \forall i, j\right\}$, we have $x_{i,j} - x_{i-1,j} =1$ for $\forall i \in \mathcal{S},\ j=1,\cdots,N$. Then, it directly follows from (\ref{5}) that $a_{i-1}=1$, $\forall i\in \mathcal{S}$ must hold.

We then prove the sufficient condition. That is, if some $\mathbf{a} \in \{0,1\}^M$ satisfies $a_{i-1} =1$, $\forall i \in \mathcal{S}$, then $\mathbf{a}$ is a feasible solution of (\ref{73}). Recall that $\varphi_{i-1}$ denotes the service type of task $(i-1)$, i.e., $u_{i-1,\varphi_{i-1}}=1$. We show that all the constraints in (\ref{53}) hold (i.e., $u_{i-1,j} a_{i-1}  \geq x_{i,j} - x_{i-1,j}, \ \forall i,j$) by separating the constraints into three non-overlapping cases: 1) $i\in \mathcal{S}$ and $j= \varphi_{i-1}$. The corresponding constraint holds because $u_{i-1,j} a_{i-1}=1$; 2) $i\in \mathcal{S}$ and $j \neq \varphi_{i-1}$. In this case, $x_{i,j} - x_{i-1,j} \leq 0$ must hold because otherwise we have $u_{i-1,j} a_{i-1} =1$, such that $\varphi_{i-1} = j$, which contradicts with our assumption that $j \neq \varphi_{i-1}$. Then, the corresponding constraint in (\ref{53}) is satisfied because $x_{i,j} - x_{i-1,j} \leq 0$; 3) $i\notin \mathcal{S}$. By the definition of $\mathcal{S}$, we infer that $x_{i,j} \leq x_{i-1,j}$ holds for all $j$, such that the corresponding constraint in (\ref{53}) holds. To sum up, all the constraints in (\ref{53}) are satisfied, such that $\mathbf{a}$ is a feasible solution of (\ref{73}). This also concludes the proof of Lemma 1. $\hfill \blacksquare$

\end{appendices}

\end{document}